\newcommand*{\ATLASLATEXPATH}{}
\pdfoutput=1
\documentclass[cernpreprint, PAPER, texlive=2016, UKenglish]{\ATLASLATEXPATH atlasdoc}
 
\usepackage[subfigure]{\ATLASLATEXPATH atlaspackage}
\usepackage{\ATLASLATEXPATH atlasbiblatex}
\usepackage{diagbox}
\usepackage{multirow}
 
\usepackage{\ATLASLATEXPATH atlasphysics}
 
\graphicspath{{logos/}{figures/}}
 
\usepackage{ANA-EXOT-2016-17-PAPER-defs}

\AtlasTitle{Search for single production of vector-like quarks
decaying into $Wb$ in $pp$ collisions at $\sqrt{s} = 13$~TeV with the ATLAS detector}
 
\author{The ATLAS Collaboration}
 
\AtlasRefCode{EXOT-2016-17}
 
\PreprintIdNumber{CERN-EP-2018-226}

\arXivId{1812.07343}
 
\HepDataRecord{84425}
 
\AtlasJournal{JHEP}
\AtlasJournalRef{JHEP 05 (2019) 164}

\AtlasDOI{10.1007/JHEP05(2019)164}
 
\AtlasAbstract{
A search for singly produced vector-like quarks $Q$, where $Q$ can be either
a $T$ quark with charge $+2/3$ or a $Y$ quark with charge $-4/3$,
is performed in proton--proton collision data at a centre-of-mass energy of 13~TeV
corresponding to an integrated luminosity of $36.1~\text{fb}^{-1}$,  recorded
with the ATLAS detector at the LHC in 2015 and 2016.
The analysis targets $Q \to Wb$ decays where the $W$ boson decays leptonically.
No significant deviation from the expected Standard Model background is observed.
Upper limits are set on the $QWb$ coupling strength and the mixing between the Standard Model sector
and a singlet $T$ quark or a $Y$ quark from a $(B,Y)$ doublet or a $(T,B,Y)$ triplet,
taking into account the interference effects with the Standard Model background.
The upper limits set on the mixing angle are as small as $|\sin{\theta_{\mathrm L}}|= 0.18$ for a singlet $T$ quark of mass 800~GeV, $|\sin{\theta_{\text{R}}}|= 0.17$ for a $Y$ quark of mass 800~GeV in a $(B,Y)$ doublet model and $|\sin{\theta_{\text{L}}}|= 0.16$ for a $Y$ quark of mass 800~GeV in a $(T,B,Y)$ triplet model.
Within a $(B,Y)$ doublet model, the limits set on the mixing parameter $|\sin{\theta_{\text{R}}}|$
are comparable with the exclusion limits from electroweak precision observables
in the mass range between about 900~GeV and 1250~GeV.
}
 
\hypersetup{pdftitle={ATLAS document},pdfauthor={The ATLAS Collaboration}}
 
\begin{document}
 
\maketitle
 
\tableofcontents
 
\section{Introduction}
\label{sec:intro}

Vector-like quarks (VLQs) are hypothetical spin-$1/2$ coloured particles with left-handed and
right-handed components that transform in the same way under the Standard Model (SM) gauge group. Therefore,
their masses are not generated by a Yukawa coupling to the Higgs boson~\cite{Aguilar-Saavedra:2013qpa}.
While the discovery of the Higgs boson ($H$) at the
Large Hadron Collider (LHC)~\cite{HIGG-2012-27,CMS-HIG-12-028}
excludes a perturbative, fourth generation of chiral quarks~\cite{Eberhardt:2012gv}, since their
contribution to loop-mediated Higgs boson couplings would significantly alter the production cross-section
and the decay rates of the Higgs boson, the effects on Higgs boson production and decay rates from
loop diagrams including VLQs are much smaller than the uncertainty in the current
measurements~\cite{Aguilar-Saavedra:2013qpa}. In many models, VLQs mix mainly with the SM quarks
of the third generation due to the large masses of the bottom and top
quarks~\cite{delAguila:1982fs,Cacciapaglia:2011fx}. Vector-like quarks appear in several extensions of the
SM that address the hierarchy problem, such as extra dimensions~\cite{Randall:1999ee},
composite Higgs~\cite{Kaplan:1983sm,Vignaroli:2012sf} and Little Higgs~\cite{Schmaltz:2005ky}
models, where they are added to the SM in multiplets. They can also appear in supersymmetric
models~\cite{Martin:2009bg} and are able to stabilise the electroweak vacuum~\cite{Xiao:2014kba}.
 
\begin{figure}[htbp]
\centering
\includegraphics[width=0.65\textwidth]{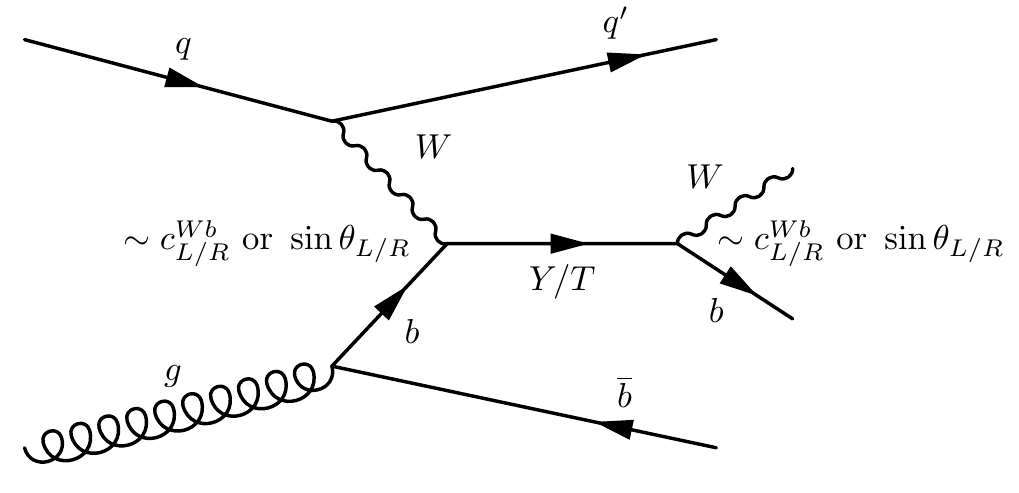}
\caption{Leading-order Feynman diagram for single $Y$/$T$ production in $Wb$ fusion and subsequent decay into $Wb$.
The production amplitude scales with $\sin \theta_{\mathrm {L, R}}$~\cite{Aguilar-Saavedra:2013qpa}
or $\cLR$~\cite{DeSimone:2012fs,Matsedonskyi:2014mna} as described in the text.}
\label{fig:FeynmanDiagram}
\end{figure}

This analysis concentrates on searches for single production of heavy vector-like quarks $Q$
produced in proton--proton ($pp$) collisions via $Wb$ fusion, $pp \rightarrow Qqb+X$, with
a subsequent decay $Q \rightarrow Wb$.
Here $Q$ can be either a $T$ quark with charge $+2/3$ or a $Y$ quark with charge $-4/3$ or
their antiquarks. An example of a leading-order Feynman diagram is presented in
Figure~\ref{fig:FeynmanDiagram}.
 
\mbox{Vector-like $T$ quarks} can belong to any weak-isospin multiplet, while $Y$ quarks cannot exist as singlets.
The interpretation used in this analysis focuses on $Y$ quarks from a $(B,Y)$ doublet or a $(T,B,Y)$
triplet, and on singlet $T$ quarks, since $T$ quarks in a $(T,B,Y)$ triplet do not couple to $Wb$
~\cite{Aguilar-Saavedra:2013qpa}.
For singlet $T$ quarks, the branching ratios ($\mathcal{B}$s) are model- and mass-dependent, but in the high-mass limit, which is considered in this analysis, they converge towards 2:1:1 ($Wb$:$Zt$:$Ht$)~\cite{Aguilar-Saavedra:2013qpa}.
Due to its charge, the $Y$ quark can decay only into $Wb$ and therefore $\mathcal{B}(Y \rightarrow Wb)=100\%$.
As a consequence, $Y$ quarks can be singly produced in $pp$ collisions only via $Wb$ fusion,
while $T$ quarks can be produced not only by $Wb$ fusion but also by $Zt$ and $Ht$ fusion.
 
Single production of vector-like quarks is enabled by their coupling to SM quarks.
As a result, searches for singly produced VLQs in $pp$ collisions can be used to probe these couplings
as a function of the VLQ mass, whereas searches for pair-produced VLQs allow limits to be set on VLQ masses; these mass limits are rather insensitive to the couplings. because the signals are produced through strong couplings. At high VLQ masses, single VLQ production can become the dominant production mechanism at the LHC,
depending on the strength of the $Qqb$ coupling. Results are presented here for two different models
that use different formulations of the Lagrangian that describes these new particles and their
interactions. In the model discussed in Ref.~\cite{Aguilar-Saavedra:2013qpa} (renormalisable theory), a mixing term
between the SM and vector-like quarks is introduced in a renormalisable extension of the SM,
while Refs.~\cite{DeSimone:2012fs,Matsedonskyi:2014mna} (non-renormalisable theory) use a phenomenological Lagrangian that is
parameterised with coupling terms, but is non-renormalisable. The main difference between these
approaches is that the Lagrangian in Refs.~\cite{DeSimone:2012fs,Matsedonskyi:2014mna} has additional
terms that allow larger production cross-sections, while the Lagrangian in Ref.~\cite{Aguilar-Saavedra:2013qpa}
gives a complete description of the dependence of the $\mathcal{B}$ on the multiplet dimension,
with left- and right-handed mixing angles $\theta_{\mathrm L}$ and $\theta_{\mathrm R}$ as model parameters.
Within a given multiplet, $\theta_{\mathrm L}$ and $\theta_{\mathrm R}$ are functionally related.
Therefore, a given value of either the left- or right-handed mixing angle fully determines
all $\mathcal{B}$s for any given heavy-quark mass.
For the interpretation in terms of coupling parameters $\cL$ and $\cR$ as introduced in
Refs.~\cite{DeSimone:2012fs,Matsedonskyi:2014mna}, assumptions must be made about the
$Q \rightarrow Wb$, $Q \rightarrow Zt$ and $Q \rightarrow Ht$ $\mathcal{B}$s in case of $Q=T$.
The relative contribution of the left- and right-handed components of the mixing and coupling
also depends on the dimension of the VLQ multiplet. For $T$ singlets, only the left-handed
component (\sinL or \cL) contributes.
For a $(B,Y)$ doublet model, results are interpreted in terms of the dominant right-handed
(\sinR) component; for a $(T,B,Y)$ triplet model, results are interpreted in terms of the
dominant left-handed (\sinL) component~\cite{Aguilar-Saavedra:2013qpa}.
The formulation of Ref.~\cite{Aguilar-Saavedra:2013qpa} also allows within a certain multiplet model a comparison of the mixing angles with constraints from electroweak precision observables, such as the ratio $R_{b}$ of the partial width for $Z \rightarrow b\bar{b}$ to the total hadronic
$Z$-boson width and the oblique parameters $S$ and $T$~\cite{Peskin:1990zt}.
A comparison of the respective Lagrangians of the renormalisable models described in Ref.~\cite{Aguilar-Saavedra:2013qpa}
and the non-renormalisable models described in Refs.~\cite{DeSimone:2012fs,Matsedonskyi:2014mna} yields a simple relation between
$\sin \theta_{\mathrm {L, R}}$ and $\cLR$: $\cLR=\sqrt{2}\sin \theta_{\mathrm{ L,R}}$ for the $T$ singlet model and $(B,Y)$ doublet model
and $\cL=2\sin \theta_{\mathrm L}$ for the $(T,B,Y)$ triplet model. This relationship
is only true within the regime of validity of the renormalisable formulation, and if one
considers only the interactions between $Q$, $W$ and $b$.
 
The ATLAS and CMS Collaborations have published searches for single and pair production of
vector-like $T$ quarks in all three decay channels~\cite{EXOT-2013-18, Aaboud:2017qpr, CMS-B2G-14-002,CMS-B2G-12-015, CMS-PAS-B2G-17-003, EXOT-2013-16, EXOT-2014-12, Aaboud:2018xuw, Sirunyan:2017usq, CMS-PAS-B2G-16-001, Aaboud:2017zfn, Sirunyan:2018omb, EXOT-2014-12, CMS-B2G-15-008, CMS-B2G-16-006, Sirunyan:2017ynj, EXOT-2013-17, CMS-PAS-B2G-16-005, EXOT-2016-35, EXOT-2017-17} and set 95\% confidence level (CL) lower limits on $T$- and $Y$-quark masses. Assuming a $\mathcal{B}$ of 100\% for the corresponding decay channel, the best observed $T$-quark mass limits are $m_{T} > 1430$~\GeV\ for $T \rightarrow Ht$~\cite{Aaboud:2018xuw}, 1340~\GeV\ for $T \rightarrow Zt$~\cite{EXOT-2016-35} and 1350~\GeV\ for $T \rightarrow Wb$~\cite{CMS-PAS-B2G-17-003}, independent of the size of the $c^{Wb}$ coupling strengths.
In Ref.~\cite{EXOT-2017-17}, seven individual analyses searching for $B\bar{B}$
or $T\bar{T}$ pair production were combined improving model-independent cross-section limits significantly over individual analyses.
$T$ quarks with a mass lower than 1310~\GeV\ are excluded for any combination of decays into SM particles by this analysis.
The observed lower limit on the pair-produced $Y$-quark mass is 1350~\GeV~\cite{Aaboud:2017zfn}.
These searches also report limits as a function of the assumed $\mathcal{B}$s.
The best observed limits are $m_{T} > 1310$~\GeV\ and $m_{T} > 1280$~\GeV\ for a weak-isospin doublet~\cite{Aaboud:2018xuw} and
singlet~\cite{Sirunyan:2018omb} respectively. Searches for single production of $T$ quarks with decays into $Zt$~\cite{EXOT-2013-17}
and single $T$/$Y$-quark production with decays into $Wb$~\cite{EXOT-2014-12} were carried out by the ATLAS Collaboration using the
Run-1 $pp$ dataset a centre-of-mass energy $\sqrt{s}$ = 8~\TeV. In the $T \rightarrow Zt$ decay channel, assuming a mixing parameter
$\sin \theta_{\mathrm{L}}$ as low as 0.7, $T$ quarks with masses between 450~\GeV\ and 650~\GeV\ are excluded~\cite{EXOT-2013-17},
while for a $QWb$ coupling strength of $\sqrt{\cTwoL+\cTwoR}=1$, the observed lower limit on the $T$-quark mass assuming
$\mathcal{B}(T\rightarrow Wb)$ = 0.5 is 950~\GeV~\cite{EXOT-2014-12}.
The CMS Collaboration studied single $T$- and $Y$-quark production using the Run-2 dataset at $\sqrt{s}$ = 13~\TeV\ collected in
2015~\cite{CMS-PAS-B2G-16-005,CMS-B2G-15-008, CMS-PAS-B2G-16-001, CMS-B2G-16-006, Sirunyan:2017ynj} and set upper limits on the
single-$T$-quark production cross-section times $\mathcal{B}(T\rightarrow Ht)$ that vary between $\mathrm{0.31~pb}$ and $\mathrm{0.93~pb}$
for $T$-quark masses in the range 1000--1800~\GeV~\cite{CMS-PAS-B2G-16-005}, as well as on the single-$T$-quark production cross-section times
$\mathcal{B}(T \rightarrow Zt)$ that vary between $\mathrm{0.98~pb}$ and 0.15~pb ($\mathrm{0.6~pb}$ and $\mathrm{0.13~pb}$)
for $T$-quark masses in the range 700--1700~\GeV\ in the right-handed (left-handed) $Tb$ ($Tt$) production channel~\cite{CMS-PAS-B2G-16-001}.
For a mass of 1000~\GeV, a $T$-quark production cross-section times branching fraction above 0.8 pb (0.7 pb) is excluded for the
$T \rightarrow Ht$ decay channel assuming left-handed (right-handed) coupling of the $T$ quark to SM particles~\cite{CMS-B2G-15-008}.
For $Y$ quarks with a coupling of 0.5 and $\mathcal{B}(Y \rightarrow Wb)$ = 1, the observed (expected) lower mass limit is
1.40 (1.0)~\TeV~\cite{CMS-B2G-16-006}.
 
This paper describes a search for $Q \rightarrow Wb$ ($Q=T$ or $Y$) production, with the prompt $W$ boson decaying
leptonically, giving a lepton + jets signature characterised by the presence of exactly
one electron or muon\footnote{Electrons and muons from decays of $\tau$-leptons from $W \rightarrow \tau \nu$ are taken into account in the selection.},
three or more jets and missing transverse momentum from the escaping neutrino. It is
assumed that $T$ quarks are produced in $Wb$ fusion only.
For single production of a $T$ quark, $Zt$ fusion could in principle contribute as well,
but can be neglected for this $T$-singlet search.
For equal values of the $TZt$ and $TWb$ couplings, the cross-section for $Zt$ fusion
is about one order of magnitude smaller than for $Wb$ fusion~\cite{Matsedonskyi:2014mna}.
For the $T$-singlet case, the $TZt$ coupling is about a factor of $\sqrt{2}$ smaller than the
$TWb$ coupling and as a result $\mathcal{B}(T \rightarrow Zt)$ is about a factor of two smaller
than $\mathcal{B}(T \rightarrow Wb)$. Since the single-VLQ production cross-section scales
with coupling squared, the $Zt$ fusion cross-section is lowered by another factor of two
compared to the $Wb$ fusion cross-section. In addition, the selection efficiency
for $tZ \rightarrow T \rightarrow Wb$ events in the search presented here is
about a factor of two smaller than for $bW \rightarrow T \rightarrow Wb$,
because in $tZ \rightarrow T \rightarrow Wb$ the accompanying top quark from the gluon
splitting leads to additional jets in the final state.
 
The analysis is optimised to search for massive VLQs with a high-momentum $b$-jet in the final state.
The $b$-jet and the charged lepton originating from the $Q$ decay are approximately back-to-back in the
transverse plane since both originate from the decay of a heavy object.
The outgoing light quark in the
process depicted in Figure~\ref{fig:FeynmanDiagram} often produces a jet in the forward region of the detector.
The second $b$-jet from the gluon splitting may be observed in either the forward or central region.
Since this $b$-jet is typically of low energy, it often falls outside the detector acceptance.
 
The main background processes with a single-lepton signature arise from top-quark pair ($t\bar{t}$) production,
single-top-quark production and $W$-boson production in association with jets ($W$+jets),
with smaller contributions from $Z$-boson production in association with jets ($Z$+jets) and from
diboson ($WW$, $WZ$, $ZZ$) production. Multijet events also contribute to the selected sample via
the misidentification of a jet or a photon as an electron or the presence of a non-prompt electron
or muon. To estimate the backgrounds from $t\bar{t}$ and $W$+jets events in a consistent and robust fashion,
two control regions (CRs) are defined. They are chosen to be orthogonal to the signal region (SR) in order to provide
independent data samples enriched in particular background sources. The reconstructed mass of the heavy-quark candidate
is used as the discriminating variable in a binned likelihood fit to test for the presence of a signal, taking into account
the interference with SM background processes.
A background-only fit to the SR and CRs is also performed to determine whether the observed event yield in the SR
is compatible with the corresponding SM background expectation. The results of the binned profile likelihood fits
are used to estimate the $Y$/$T$ $\cLR$ coupling limits. In the case of the right-handed $Y$ quark in a $(Y,B)$ doublet model,
where the interference effect with the SM is much smaller than for the other models under consideration,
a limit on the production cross-section is also quoted.
 
\section{ATLAS detector}
\label{sec:detector}
The ATLAS detector~\cite{PERF-2007-01} at the LHC is a multipurpose particle detector with a forward--backward symmetric cylindrical geometry that covers nearly the entire solid angle around the collision point.\footnote{ATLAS uses a right-handed coordinate system with its origin at the nominal interaction point (IP) in the centre of the detector. The positive $x$-axis is defined by the direction from the IP to the centre of the LHC ring, with the positive $y$-axis pointing upwards, while the beam direction defines the $z$-axis. Cylindrical coordinates $(r,\phi)$ are used in the transverse plane, $\phi$ being the azimuthal angle around the $z$-axis. The pseudorapidity $\eta$ is defined in terms of the polar angle $\theta$ by $\eta=-\ln\tan(\theta/2)$. The transverse momentum ($p_{\mathrm{T}}$) is defined relative to the beam axis and is calculated as $p_{\mathrm{T}} = p \sin(\theta)$.}
It consists of an inner tracking detector (ID) surrounded by a thin superconducting
solenoid magnet producing an axial 2~T magnetic field, fine-granularity electromagnetic (EM) and hadronic calorimeters, and a muon spectrometer (MS) incorporating three large air-core toroid magnet assemblies. The ID consists of a high-granularity silicon pixel detector,
including an insertable B-layer~\cite{capeans:1291633,Abbott:2018ikt} added in 2014, and a silicon microstrip tracker, together providing
charged-particle tracking information in the pseudorapidity region $|\eta|<2.5$.
It is surrounded by a transition radiation tracker, which enhances electron identification information in the region $|\eta|<2.0$.
The EM calorimeter is a lead/liquid-argon sampling detector, divided into a barrel region ($|\eta|<1.475$) and two endcap regions ($1.375<|\eta|<3.2$), which provides energy measurements of electromagnetic showers.
Hadron calorimetry is also based on the sampling technique, with either scintillator tiles or liquid argon as the active medium and with steel, copper, or tungsten as the absorber material. The calorimeters cover the region $|\eta|<4.9$.
The MS measures the deflection of muons within $|\eta|<2.7$ using three layers of high-precision tracking chambers located in a toroidal field of approximately 0.5~T and 1~T in the central and endcap regions
respectively. The MS is also instrumented with separate trigger chambers covering $|\eta|<2.4$. A two-level trigger system~\cite{TRIG-2016-01}, using custom hardware followed by a software-based level, is used to reduce the trigger rate to a maximum of around 1 kHz for offline data storage.

 
\section{Physics object reconstruction}
\label{sec:selection}
The data used in this search correspond to an integrated luminosity of $36.1~\text{fb}^{-1}$ of $pp$ collisions at a centre-of-mass energy of $\sqrt{s}$ = 13~\TeV\ recorded in 2015 and 2016 with
the ATLAS detector. Only data-taking periods with stable beam collisions and all relevant ATLAS detector components functioning normally are considered. In this dataset, the average number of simultaneous $pp$ interactions per bunch crossing, or `pile-up', is approximately 24.
 
The final states considered in this search require the presence of one charged lepton (electron or muon) candidate and multiple hadronic jets.
Single-electron and single-muon triggers with low transverse-momentum (\pt) thresholds and lepton isolation requirements (in 2016 only) are combined in a logical OR with higher-threshold triggers without any isolation requirements to give maximum efficiency. For electrons, triggers with a \pt threshold of 24 (26)~\GeV\ in 2015 (2016) and isolation requirements (in 2016 only) are used along with triggers with a $60~\gev$ threshold and no isolation requirement, and with a 120 (140)~\GeV\ threshold with looser identification criteria. For muons, triggers with \pt thresholds of 20 (26)~\GeV\ and isolation requirements in 2015 (2016) are combined with a trigger that has a \pt threshold of $50~\gev$ and no isolation requirements in both years. In addition, events must have at least one reconstructed vertex with two or more tracks with \pt above 0.4~\GeV\ that is consistent with the beam-collision region in the  $x$--$y$ plane.
If multiple vertices are reconstructed, the vertex with the largest sum of the squared \pt of its associated tracks is taken as the primary vertex. For the final states considered in this analysis, the
vertex reconstruction and selection efficiency is close to 100\%.
 
Electron candidates~\cite{PERF-2013-05,ATLAS-CONF-2016-024,ATL-PHYS-PUB-2016-015} are reconstructed from isolated energy deposits (clusters) in the EM calorimeter, each matched to a reconstructed ID track,
within the fiducial region of $ | \eta_{\text{cluster}} | < 2.47 $, where $\eta_{\text{cluster}}$ is the pseudorapidity of the centroid of the calorimeter energy deposit associated
with the electron candidate. A veto is placed on electrons in the transition region between the barrel and endcap electromagnetic calorimeters, $1.37 <|\eta_{\text {cluster}}| < 1.52$.
Electrons must satisfy the tight likelihood identification criterion, based on shower-shape and track--cluster matching variables, and must have a transverse energy
$\ET = E_\mathrm{cluster}/\cosh(\eta_\mathrm{track})> \SI{25}{\GeV}$, where $E_\mathrm{cluster}$ is the electromagnetic cluster energy and $\eta_\mathrm{track}$ the track pseudorapidity.
Muons are reconstructed~\cite{PERF-2015-10} by combining a track reconstructed in the ID with one in the MS,
using the complete track information from both detectors and accounting for the effects of energy loss and multiple scattering in the material of the detector structure.
The muon candidates must satisfy the medium selection criteria~\cite{PERF-2015-10} and are required to be in the central region of $|\eta|<2.5$. To reduce the contribution of leptons from hadronic decays (non-prompt leptons), electrons and muons must satisfy isolation criteria that include both track and calorimeter information, and are  tuned to give an overall efficiency of 98\%, independent of the $\pT$ of the lepton. Electron and muon candidates are required to be isolated from additional tracks within a cone around their directions with a radius of ${\Delta}R \equiv \sqrt{(\Delta\eta)^{2} + (\Delta\phi)^{2}}$ with ${\Delta}R =
\min(0.2, 10~\GeV{}/p_{\text T})$~\cite{ATLAS-CONF-2016-024} for electrons and ${\Delta}R = \min(0.3, 10~\GeV{}/p_{\text T})$ for muons~\cite{PERF-2015-10}.
The lepton calorimeter-based isolation variable is defined as the sum of the calorimeter transverse energy deposits in a cone of size ${\Delta}R =0.2$, after subtracting the contribution
from the energy deposit of the lepton itself and correcting for pile-up effects, divided by the lepton $p_{\text T}$.
The significance of the transverse impact parameter $d_{0}$, calculated relative to the measured beam-line position, is required to satisfy $|d_{0} / \sigma(d_{0})|< 5 $ for electrons and $|d_{0} / \sigma(d_{0})|< 3 $ for muons, where $\sigma(d_{0})$ is the uncertainty in $d_{0}$. Finally, the lepton tracks are matched to the primary vertex of the event by requiring the longitudinal impact parameter $z_{0}$
to satisfy $|z_{0}\sin\theta_{\mathrm{track}}|< 0.5$~mm, where $\theta_{\mathrm{track}}$ is the polar angle of the track.\footnote{The longitudinal impact parameter $z_{0}$
is the difference between the longitudinal position of the track along the beam line at the point where the transverse impact parameter ($d_{0}$) is measured and the longitudinal
position of the primary vertex.}
 
The leptons satisfying the criteria described above are used in the selection of events in the signal and control regions. The estimation of background from non-prompt and fake leptons with the Matrix Method~\cite{ATLAS-CONF-2014-058}, described in Section~\ref{sec:fake}, uses `loose' leptons in addition to the above `tight' leptons, where the tight sample is a subset of the loose sample. The `loose' selection requires that the muon (electron) satisfies the medium (likelihood medium) requirements, but does not need to satisfy isolation criteria as defined in Refs.~\cite{ATLAS-CONF-2016-024, PERF-2015-10}.
 
Jets are reconstructed from three-dimensional topological calorimeter energy clusters~\cite{PERF-2014-07} using the anti-$\mathit{k_{t}}$ algorithm~\cite{Cacciari:2008gp, Cacciari:2011ma} with a radius parameter of $0.4$~\cite{PERF-2011-03}. Each topological cluster is calibrated to the electromagnetic energy scale prior to jet reconstruction. The reconstructed jets from the clusters are then calibrated to the particle level by the application of corrections derived from simulation and from dedicated calibration samples of $pp$ collision data at $\sqrt{s}$ = 13~\TeV~\cite{ATLAS-CONF-2015-037, Aaboud:2017jcu}. Data quality criteria are imposed to identify jets arising from non-collision sources or detector noise, and any event containing at least one such jet is removed~\cite{ATLAS-CONF-2015-029}.
Finally, jets considered in this analysis are required to have $\pT > 25$~\GeV. The pseudorapidity acceptance for jets differs between
different selections: central jets are required to have $|\eta| < 2.5$, while forward jets are defined to have $2.5 < |\eta|< 4.5$.
Furthermore, jets with a $\pT < 60$~\GeV\ and $|\eta| < 2.4$ are required to satisfy criteria implemented in the jet vertex tagger algorithm~\cite{PERF-2014-03} designed to select jets that originate from the hard scattering and reduce the effect of in-time pile-up.
 
The identification of jets from $b$-quark decays ($b$-tagging) is beneficial in this analysis.
To identify (tag) jets containing $b$-hadrons (henceforth referred to as $b$-jets), a multivariate
discriminant is used that combines information about the impact parameters of inner-detector tracks associated with the jet,
the presence of displaced secondary vertices, and the reconstructed flight paths of $b$- and $c$-hadrons inside the jet~\cite{ATL-PHYS-PUB-2015-039,ATLAS-CONF-2014-046, PERF-2012-04, PERF-2016-05}.
Jets are considered to be $b$-tagged if the value of the multivariate discriminant is larger than a certain threshold.
The criterion in use is only calculated for central jets ($|\eta|< 2.5$) with $\pt > \SI{25}{\GeV}$ and has an efficiency of approximately 85\% for $b$-jets in simulated \ttbar\ events.
The rejection factor against jets originating from light quarks and gluons (henceforth referred to as light-flavour jets)
is about 34, and that against jets originating from charm quarks ($c$-jets) is about 3~\cite{PERF-2012-04}, determined
in simulated \ttbar\ events. Correction factors are defined to correct the tagging rates in the simulation to match the efficiencies measured in the data control samples~\cite{PERF-2012-04, Aaboud:2018xwy}.
 
To avoid counting a single detector response as two objects, an overlap removal procedure is used.
Jets overlapping with identified electron
candidates within a cone of $\Delta R = 0.2$ are removed, as the jet and the
electron are very likely to be the same physics object. If the nearest jet surviving this requirement
is within $\Delta R = 0.4$ of an electron, the electron is discarded,
to ensure it is sufficiently separated from nearby jet activity.
Muons are removed if they are separated from the nearest jet by $\Delta R < 0.4$,
to reduce the background from muons from heavy-flavour hadron decays inside jets.
However, if this jet has fewer than three associated tracks,
the muon is kept and the jet is removed instead;
this avoids an inefficiency for high-energy muons undergoing significant energy loss in the calorimeter.
 
The missing transverse momentum $\vec{E}_{\text{T}}^{\text{miss}}$ (with magnitude \MET) is a measure
of the momentum of the escaping neutrinos. It is defined as the negative vector sum  of the transverse
momenta of all selected and calibrated objects (electrons, muons, photons, hadronically decaying
$\tau$-leptons and jets) in the event, including a term to account for energy from soft particles
which are not associated with any of the selected objects~\cite{PERF-2016-07}. This soft term is
calculated from inner-detector tracks matched to the selected primary vertex to make it resilient
to contamination from pile-up interactions~\cite{PERF-2016-07}.
 
\section{Background and signal modelling}
\label{sec:modelling}
Monte Carlo (MC) simulation samples are used to model the expected signal and SM background distributions.
The MC samples were processed either through the full ATLAS detector simulation \cite{SOFT-2010-01} based
on {\scshape Geant4} \cite{agostinelli:2002hh} or through a faster simulation making
use of parameterised showers in the calorimeters~\cite{ATL-PHYS-PUB-2010-013}.
Effects of both in-time and out-of-time pile-up, from additional $pp$ collisions in the same and nearby bunch crossings,
were modelled by overlaying minimum-bias interactions generated with
{\PYTHIA} 8.186~\cite{Sjostrand:2007gs} according to the luminosity profile of the recorded data.
The distribution of the number of additional $pp$ interactions in the MC samples was reweighted to
match the pile-up conditions observed in data.
All simulated samples used {\scshape EvtGen}~\cite{lange:2001uf} to model the decays of heavy-flavour hadrons,
except for processes modelled using the {\SHERPA} generator~\cite{Gleisberg:2008ta}.
All simulated events were processed using the same reconstruction algorithms and analysis selection requirements
as for the data, but small corrections, obtained from comparisons of simulated events with data in dedicated
control regions, were applied to trigger and object reconstruction efficiencies, as well as detector resolutions,
to better model the observed response. The main parameters of the MC samples used in this search are summarised
in Table~\ref{table:MCsamples}. Samples for all SM background processes were generated with the full {\scshape Geant4}
model of the ATLAS detector.

\subsection{Background modelling}
\label{sec:bkgmodelling}
Top-quark pair events were generated with the next-to-leading-order (NLO) generator {\PowhegBox} 2.0~\cite{nason:2004rx,alioli:2010xd,Frixione_test_2007} using the {\textsc CT10} parton distribution function (PDF) set~\cite{Lai:2010vv}, interfaced to {\PYTHIA} 6.428~\cite{Sjostrand:2006za} with the CTEQ6L PDF set~\cite{Pumplin:2002vw} and the {\textsc Perugia 2012} ({\textsc P2012}) set of tuned parameters for the underlying event (UE)~\cite{Skands:2010ak}.
The hard-process factorisation scale $\muF$ and renormalisation scale $\muR$ were set to the default {\PowhegBox} values $\mu = (m^{2}_{t} + p^{2}_{\text{T,top}})^{1/2} $, where $m_t$  is the top-quark mass, $m_t = 172.5~\gev$, and
$p_{\text{T,top}}$ is the top-quark transverse momentum evaluated for the underlying Born configuration.
The $h_{\text {damp}}$ parameter, which controls the transverse momentum of the first
additional gluon emission beyond the Born configuration, is set equal to the
mass of the top quark. The main effect of this choice is to regulate the
high-$p_T$ emission against which the $t\bar{t}$ system recoils.
The sample was generated assuming that the top quark decays exclusively through $t \to Wb$.
 
Alternative $\ttbar$ samples were produced to model uncertainties in this process.
The effects of initial- and final-state radiation (ISR/FSR) were explored using two alternative {\PowhegBox} 2.0 + {\Pythia} 6.428 samples: one with $h_{\text {damp}}$ set to $2 m_t$, the renormalisation and factorisation scales
set to half the nominal value and using the {\textsc P2012} high-variation UE tuned parameters, giving more radiation, and
another with {\textsc P2012} low-variation UE
tuned parameters, $h_{\text {damp}}=m_t$ and the renormalisation and factorisation scales set to twice the nominal value, giving less radiation~\cite{ATL-PHYS-PUB-2015-011}.
The values of $\muR$, $\muF$ and $h_{\text {damp}}$ were varied together because
these two variations were found to cover the full set of uncertainties obtained by changing the scales and the $h_{\text {damp}}$ parameter independently. To provide a comparison with a different parton-shower model, an additional \ttbar\ sample was generated using the same {\PowhegBox} settings as the nominal {\PowhegBox} 2.0 + {\PYTHIA} 6.428 sample,
but with parton showering, hadronisation, and the UE simulated with {\Herwig}++ 2.7.1~\cite{Bahr:2008pv} with the {\textsc UEEE5} tuned parameters~\cite{Seymour:2013qka} and the corresponding CTEQ6L1 PDF set.
Additional $\ttbar$ simulation samples were generated using {\scshape Madgraph5}\_a{\scshape MC@NLO} 2.2.1~\cite{Alwall:2014hca} interfaced to {\Herwig}++ 2.7.1 to determine the systematic uncertainties related to the use of different models for the hard-scattering generation, while maintaining the same parton shower model.
 
The $\ttbar$ prediction was normalised to the theoretical cross-section for the inclusive \ttbar\ process of $832^{+46}_{-51}$~pb obtained with {\scshape Top++}~\cite{Czakon:2011xx},
calculated at next-to-next-to-leading order (NNLO) in QCD and including resummation of next-to-next-to-leading logarithmic (NNLL) soft gluon terms~\cite{xs1,xs2,xs3,xs4,xs5}.
Theoretical uncertainties result from variations of the factorisation and renormalisation scales, as well as from uncertainties in the PDF and strong coupling constant $\alpha_{\text S}$. The latter two represent the largest contribution to the overall theoretical uncertainty
in the cross-section and are calculated using the PDF4LHC prescription~\cite{Butterworth:2015oua}.
 
Single-top-quark background processes corresponding to the $Wt$ and $s$-channel production mechanisms were generated with {\PowhegBox} 1.0 at NLO~\cite{Frederix:2012dh} using the {\textsc CT10} PDF set.
Overlaps between the \ttbar\ and $Wt$ final states were removed using the ``diagram removal'' scheme (DR)~\cite{Frixione_test_2005vw, Frixione:2008yi}.
The ``diagram subtraction'' scheme (DS)~\cite{Frixione:2008yi} was considered as an alternative method,
and the full difference between the two methods assigned as a shape and normalisation uncertainty~\cite{White:2009yt}.
Events from $t$-channel single-top-quark production were generated using the {\PowhegBox} 1.0~\cite{Frederix:2012dh} NLO generator, which uses the four-flavour scheme.
The fixed four-flavour PDF set \textsc{CT10}f\textsc{4} was used for the matrix-element calculations.
All single-top-quark samples were normalised to the approximate NNLO theoretical cross-sections~\cite{Kidonakis:2011wy,Kidonakis:2010ux,Kidonakis:2010tc}. {\PYTHIA} 6.428 with the {\textsc P2012} set of tuned parameters was used to model the parton shower, hadronisation and underlying event.
Additional single-top-quark samples were generated using the same {\PowhegBox} settings as the nominal sample, while parton showering, hadronisation, and the UE were simulated with {\Herwig}++ 2.7.1. The ISR/FSR effects were
explored using alternative {\PowhegBox} 2.0 + {\Pythia} 6.428 samples with a set of {\textsc P2012} high- and low-variation UE tuned parameters. Another set of single-top-quark samples was generated using {\scshape Madgraph5}\_a{\scshape MC@NLO} 2.2.1 interfaced to {\Herwig}++ 2.7.1 to determine the systematic uncertainties associated with the choice of NLO generator.
 
Samples of $W/Z$+jets events were generated with the {\scshape Sherpa} 2.2.0 generator.
The matrix-element calculation was performed with up to two partons at NLO and up to four partons at leading order (LO) using
{\scshape Comix}~\cite{Gleisberg:2008fv} and {\scshape OpenLoops}~\cite{Cascioli:2011va}. The matrix-element calculation
was merged with the {\scshape Sherpa} parton shower~\cite{Schumann:2007mg} using the ME+PS@NLO
prescription~\cite{Hoeche:2012yf}. The PDF set used for the matrix-element calculation was {\textsc CT10} with a
dedicated parton shower tuning developed by the {\scshape Sherpa} authors.
The $W$+jets and $Z$+jets samples were normalised to the NNLO theoretical cross-sections for inclusive $W$ and $Z$ production calculated with {\scshape FEWZ}~\cite{Anastasiou:2003ds}. Samples generated with
{\scshape Madgraph5}\_a{\scshape MC@NLO} 2.2.1+ {\Pythia} 8.186 were compared with the nominal $W$+jets samples to determine the systematic uncertainties associated with the choice of generator.
 
Diboson events ($WW/WZ/ZZ$+jets) with one of the bosons decaying hadronically and the other leptonically
were generated with the NLO generator {\scshape Sherpa} 2.1.1 and include processes containing up to four electroweak vertices.
The matrix element included  up to one ($ZZ$) or zero ($WW$, $WZ$) additional partons at NLO and up to three partons at LO
using the same procedure as for $W/Z$+jets.
All diboson samples were normalised to their NLO theoretical cross-sections provided by {\scshape Sherpa}.
Processes producing smaller backgrounds are also considered, and include $\ttbar$$V$ ($V$ = $W$,$Z$) and $\ttbar$$H$. The $\ttbar$$V$ processes were simulated with {\scshape Madgraph5}\_a{\scshape MC@NLO} generator using
the NNPDF2.3 PDF set, interfaced to \Pythia8~\cite{Sjostrand:2014zea} with the A14 UE tune. The $\ttbar$$H$ process was modelled using {\scshape Madgraph5}\_a{\scshape MC@NLO} interfaced to {\Herwig}++ 2.7.1.

\begin{table}[ht]
\centering
\caption{\label{table:MCsamples}
Generators used to model the signals and different background processes. The parameter tune for the underlying event,
PDF set, and the highest-order perturbative QCD (pQCD) accuracy used for the normalisation of each sample is given.
All processes, except for $Yqb$ signals, were generated at NLO in QCD. The LO cross-sections calculated for the $Yqb$ signal
processes in the simulation were normalised to the NLO theoretical cross-section taken from Ref.~\cite{Matsedonskyi:2014mna}.}
{\small
\begin{tabular}{lllll}
\toprule\toprule
Process &    Generator  		 & Tuned parameters & PDF set  &  Inclusive cross-section  \\
& + parton showering/hadronisation  &	&	   &  order in pQCD\\
\midrule
{\textbf {$Yqb$}} & {\scshape Madgraph5}\_a{\scshape MC@NLO} 2.2.3 &{\textsc A14}  & {\textsc NNPDF2.3} & NLO \\
& + \PYTHIA 8.210 & &  & \\
\\
\midrule
{\textbf \ttbar} & {\PowhegBox} 2.0
& {\textsc P2012} & {\textsc CT10} & NNLO+NNLL \\
& + \PYTHIA 6.428 & & & \\
\\
\midrule
{\textbf {Single top}} & {\PowhegBox} 1.0
& {\textsc P2012} & {\textsc CT10} & NNLO+NNLL\\
& + \PYTHIA 6.428 & & & \\

\\
\midrule
{\textbf {Dibosons}}  & \SHERPA 2.1.1 & Default & {\textsc CT10} & NLO \\
$WW$, $WZ$, $ZZ$ & & & & \\
\\
\midrule
{\textbf {$W$/${Z}$ + jets}} & \SHERPA 2.2.0 & Default & {\textsc CT10} & NNLO  \\
\midrule
{\textbf \ttbar $V$} &{\scshape Madgraph5}\_a{\scshape MC@NLO} 2.2.3 &{\textsc A14}  & {\textsc NNPDF2.3} &NLO  \\
& + \PYTHIA 8.210 & &  & \\
\midrule
{\textbf \ttbar $H$} &{\scshape Madgraph5}\_a{\scshape MC@NLO} 2.2.3 &{\textsc CTEQ6L1}  & {\textsc CT10} &NLO  \\
& + {\Herwig}++ 2.7.1 & &  & \\
 
\bottomrule \bottomrule
\end{tabular}
}
\end{table}

\subsection{Signal modelling}
\label{sec:Signal}
Simulated events for signal processes were generated at LO in the four-flavour
scheme with the {\scshape Madgraph5}\_a{\scshape MC@NLO} 2.2.3 generator using
the NNPDF2.3 PDF set, interfaced to \Pythia8 for parton showering and hadronisation.
Samples of $Yqb$ signals were produced for masses ranging from 800~\GeV\ to 2000~\GeV\ in steps
of 100~\GeV\ with equal left-handed and right-handed coupling strengths of $\kappa_{T} = 0.5$~\cite{Buchkremer:2013bha}.
The coupling parameter $\kappa_{T}$ in the model described in Ref.~\cite{Buchkremer:2013bha} used
for the signal production is related to the coupling parameters $c_{\mathrm {L,R}}^{Wb}$ in
Ref.~\cite{Matsedonskyi:2014mna}
via $\kappa_T f(m)  = c_{\mathrm{L,R}}^{Wb}/\sqrt{2}$, where $f(m)\approx \sqrt{1/(1+\mathcal{O}(m_{Q}^{-4}))}$ with
$m_{Q}$ the VLQ mass in \GeV, and therefore $\kappa_T \approx c_{\mathrm{L,R}}^{Wb}/\sqrt{2}$ to a very good approximation.
These samples were processed either through the full detector simulation or through the faster simulation.
The normalisation of signal events produced with the faster simulation was scaled up by 7.2$\%$ to correct for efficiency
differences.
 
Since the kinematic distributions of the decay products for the $T$ quark and $Y$ quark in the $Wb$ decay channel
are the same, only $Y$ signal samples were generated and they were used to derive the results also for the $Tqb$ signals.
Other possible decay modes of the $T$ quark ($T \rightarrow Zt$, $T \rightarrow Ht$) have negligible acceptance
in this search.
The kinematics of the final-state particles are very similar for left-handed and right-handed couplings,
and hence the acceptances for the two chiralities are found to be equal. The LO cross-sections calculated
for the signal processes in the simulation were normalised to the next-to-leading-order
benchmark calculation from Ref.~\cite{Matsedonskyi:2014mna}, which is performed
in the narrow-width approximation (NWA).
The single-VLQ production cross-sections and the decay widths of the VLQ resonances are mass- and
coupling-dependent. The VLQ width increases with increasing mass and coupling values such that,
for sufficiently large masses and couplings, the NWA is no longer valid.
The ratio of the single-VLQ production cross-section without the NWA to that with the NWA, calculated at LO
using {\scshape Madgraph5}\_a{\scshape MC@NLO} 2.2.3, was used to correct the NLO cross-section from
Ref.~\cite{Matsedonskyi:2014mna} as function of VLQ mass and coupling.
 
\begin{figure}[ht]
\centering
\subfigure[]{
\includegraphics[width=0.4\textwidth]{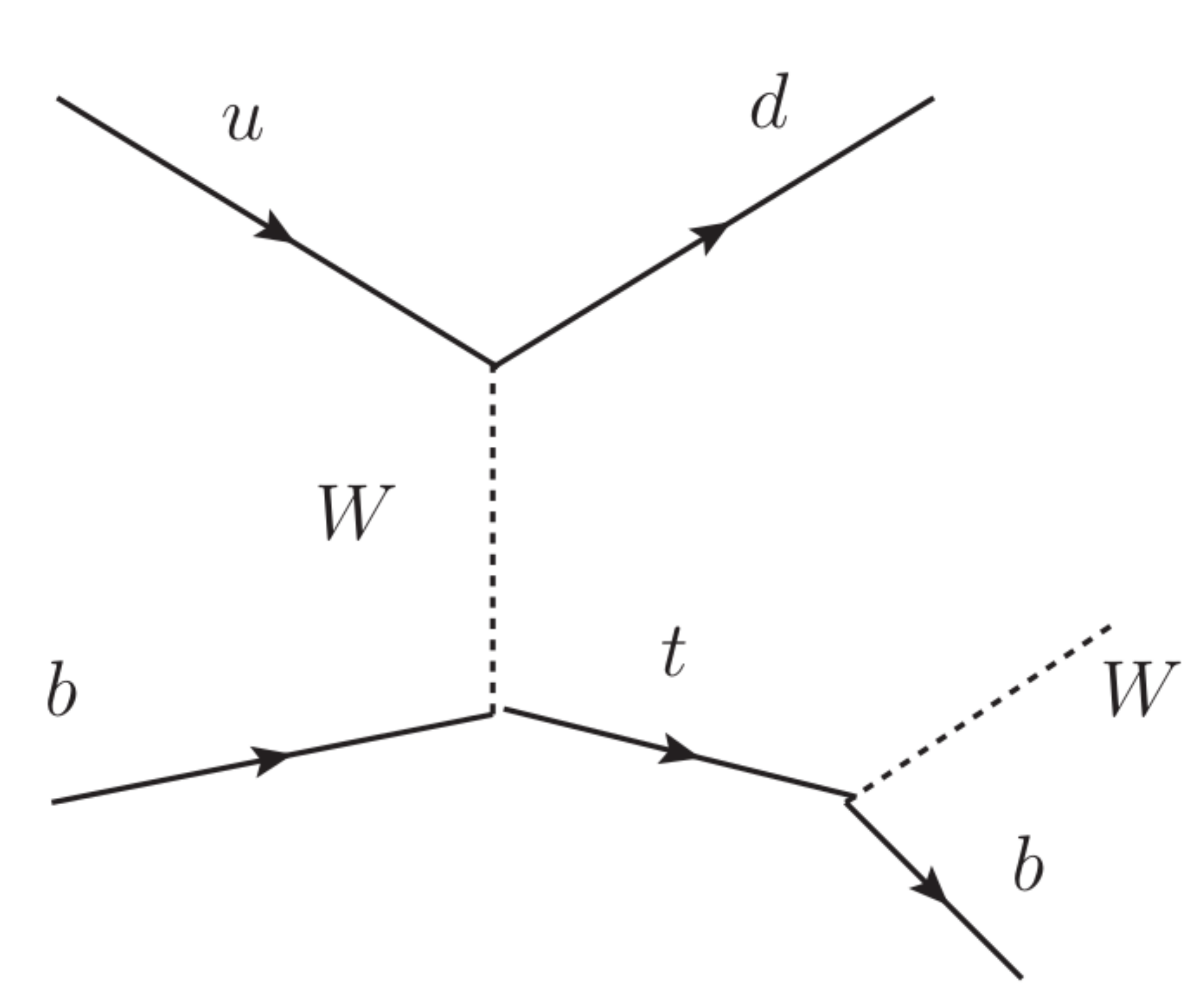}
\label{Fig:interferences_with_T_signal}
}
\hfill
\centering
\subfigure[]{
\includegraphics[width=0.35\textwidth]{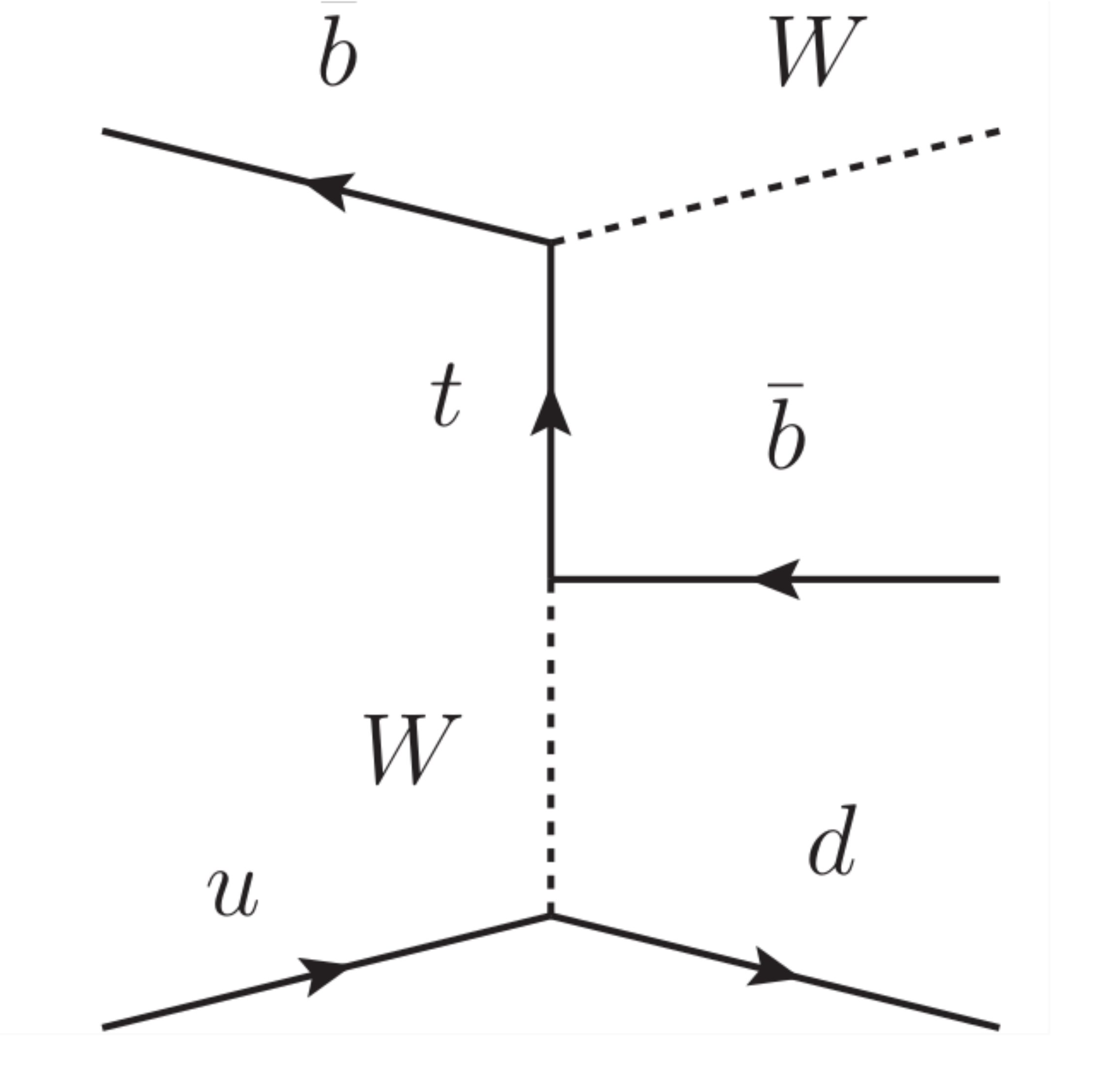}
\label{Fig:interferences_with_Y_signal}
}
\null
\caption{Leading-order Feynman diagrams for the SM processes that interfere with $T$-quark or $Y$-quark production, respectively, as described in the text: \subref{Fig:interferences_with_T_signal} $t$-channel single-top-quark production where the top quark is far off-shell and \subref{Fig:interferences_with_Y_signal} electroweak $W^{-}bq$
production. }
\label{Fig:interferences_with_signal}
\end{figure}

Sizeable interference effects between the amplitude for VLQ signal production and the SM are possible.
In the analysis, two scenarios are considered:
\begin{enumerate}
\item $T$-quark production in a $T$ singlet model, in which the $T$ quark has only a left-handed coupling~\cite{Aguilar-Saavedra:2013qpa}.
The SM process that interferes in this case is $t$-channel single-top-quark production where the top quark is far off-shell as illustrated in Figure~\ref{Fig:interferences_with_T_signal}.
\item $Y$-quark production in a $(T,B,Y)$ triplet or $(B,Y)$ doublet model, in which the $Y$ quark has only a left-handed coupling or right-handed coupling. The SM process that interferes with $Y$-quark
production is electroweak $W^{-}bq$ production\footnote{The charge-conjugated state $W^{+}\bar{b}q$ interferes with the $\bar{Y}$ quark.} as shown in Figure~\ref{Fig:interferences_with_Y_signal}. 
Two cases are considered: a) the $Y$ quark has only a left-handed coupling, which is realised e.g.\ in a $(T,B,Y)$ triplet model, in which the right-handed coupling is heavily suppressed~\cite{Aguilar-Saavedra:2013qpa}.
Since in  the $(T,B,Y)$ triplet model the $T$ quark does not couple to $Wb$, $T$-quark production
does not contribute to the final state under consideration; b) the $Y$ quark has only a right-handed coupling, which is realised e.g.\ in a $(B,Y)$ doublet model, in which the left-handed coupling is heavily suppressed. The interference effect for the $Y$ quark with a right-handed coupling is much smaller than that for the $Y$ quark with a left-handed coupling.
\end{enumerate}
These SM contributions (i. e. $\sigma_{\textrm{SM}}$) were not modelled in the ATLAS MC simulations.

In order to determine the signal yield and acceptance for different signal couplings, the samples of simulated signal
events produced with the nominal coupling strength of $\kappa_T$ = 0.5 are corrected on an event-by-event basis
using reweighting factors. These factors are obtained by comparing the target VLQ mass distribution
in generated signal samples, at particle level, with the nominal one. The reweighting takes three effects into account:
1) the effect of interference calculated at LO, 2) the change in cross-section when going from
LO to NLO, 3) the effect from the variation of the coupling strength.
The method is validated with fully reconstructed signal samples with varied coupling strengths.
The matrix-element squared for the process $pp \rightarrow Wbq$ is given by
\begin{linenomath*}
\begin{equation*}
|M|^2 = |M_{\textrm{SM}}|^2 + |M_{\textrm{VLQ}}|^2 + 2 Re(M_{\textrm{SM}}^{*} M_{\textrm{VLQ}}).
\end{equation*}
\end{linenomath*}
As a result, the total cross-section for $pp \rightarrow Wb q$ at LO can be written as
$\sigma_{\textrm{tot}}^{\textrm{LO}} = \sigma_{\textrm{SM}}^{\textrm{LO}} + \sigma_{\textrm{VLQ}}^{\textrm{LO}} + \sigma_{\textrm{I}}^{\textrm{LO}}$
with the LO SM cross-section $\sigma_{\textrm{SM}}^{\textrm{LO}}$, the LO VLQ cross-section $\sigma_{\textrm{VLQ}}^{\textrm{LO}}$
and the interference-term cross-section $ \sigma_{\textrm{I}}^{\textrm{LO}}$. Since the $K$-factor quantifying the ratio
between NLO and LO cross-sections is significantly larger than one for VLQ production, the interference
effect has to be modelled at NLO. This modelling uses the $K$-factors for SM production, $K_{\text{SM}}$,
and for VLQ production, $K_{\text{VLQ}}$, writing the total cross-section for $pp \rightarrow W b q$
at NLO as
\begin{linenomath*}
\begin{equation}
\sigma_{\textrm{tot}}^{\mathrm{NLO}} = K_{\textrm{SM}} \sigma_{\textrm{SM}}^{\mathrm{LO}}
+ K_{\textrm{VLQ}} \sigma_{\textrm{VLQ}}^{\mathrm{LO}}
+ \sqrt{K_{\textrm{SM}} \cdot K_{\textrm{VLQ}}} \sigma_{\mathrm{I}}^{\mathrm{LO}}.
\label{eq:xsec}
\end{equation}
\end{linenomath*}
The $K_{\text{VLQ}}$ values as a function of the VLQ mass are taken from Ref.~\cite{Matsedonskyi:2014mna}.
There is no dedicated NLO calculation available for the $K_{\text{SM}}$ factor for $t$-channel single-top-quark production
with $t$-quarks far off-shell. This $K_{\text{SM}}$ factor is set to unity since the $K$-factor for
$t$-channel single-top-quark production for on-shell $t$-quarks is very close to one~\cite{Berger:2016oht}.
Since there is no dedicated NLO calculation in the literature for electroweak SM $W^{-}bq$ production interfering
with the $Y$ production amplitude, $K_{\text{SM}}$ is set to unity in this case as well.
No systematic uncertainties are assigned to any of the
$K_{\text{VLQ}}$ or $K_{\text{SM}}$ factors, because it is assumed that they correspond to the particular model assumptions.
To obtain the reweighting factors $r$, events were generated at LO using {\scshape Madgraph5}\_a{\scshape MC@NLO} 2.2.3 and
$r$ calculated as \begin{linenomath*}\begin{equation}
r(m_{Wb};c,c_{0}) =\frac{K_{\textrm{VLQ}} f_{\textrm{VLQ}}(m_{Wb};c) + \sqrt{K_{\text{SM}} \cdot K_{\textrm{VLQ}}} f_{\mathrm{I}}(m_{Wb};c)}{f_{\textrm{VLQ}}(m_{Wb};c_{0})},
\label{eq:ratio}\end{equation}\end{linenomath*}
where $c_{0}$ is the nominal coupling used in the simulation, $c$ is the coupling value of interest,
and the functions $f_{\textrm{VLQ}}(m_{Wb};c)$ and $f_{\mathrm{I}}(m_{Wb};c)$ describe the $Wb$ invariant mass distributions
at particle level scaled to the LO cross-sections $\sigma_{\textrm{SM}}^{\mathrm{LO}}$ and $\sigma_{\mathrm{I}}^{\mathrm{LO}}$ respectively.
The reweighting assumes that the phase change as a function of $m_{Wb}$ for the VLQ and SM amplitudes at NLO is the same as at LO.
 
\begin{figure}
\centering
\includegraphics[width=0.7\textwidth]{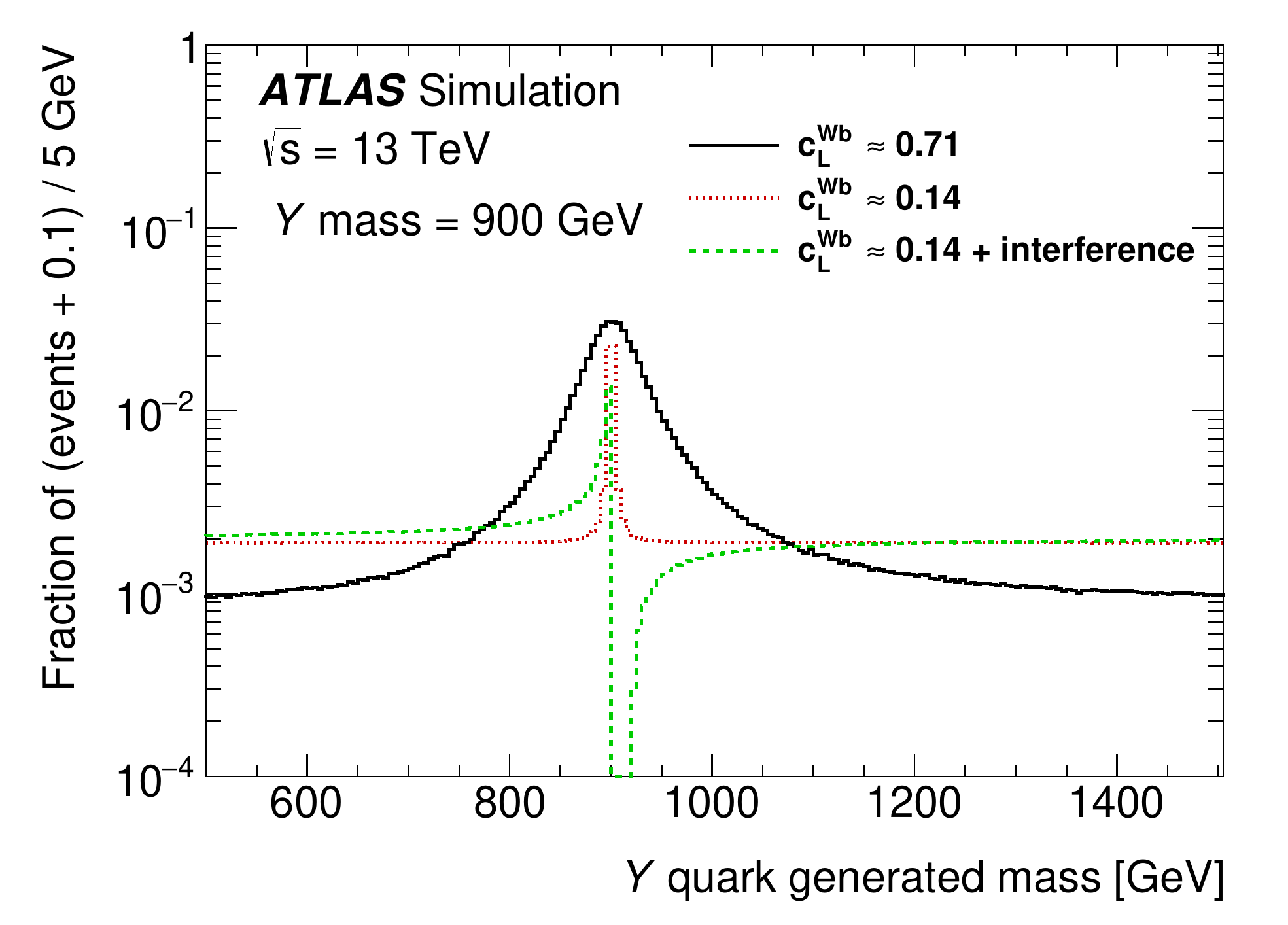}
 
\caption{The generated mass distributions at particle level for a $Y$ quark with a mass of 900~\GeV,
for a coupling strength of $c_{0}=\kappa_{T}\approx 0.5$ and $\cL\approx 1/\sqrt{2}$ ($\cR=0$ , solid line)
and of $c_0=\cL = 0.14$ (dotted line) as defined in Ref.~\cite{Buchkremer:2013bha}. The distribution for a right-handed only and left-handed only $Y$ quark (solid line) is the same. The dashed line shows the generated vector-like quark mass distribution at particle level of a left-handed $Y$ signal with a mass of 900~\GeV, coupling strength of \mbox{$\cL = 0.14$} and interference effects with
the SM included.
The interference effects lead to negative entries in some bins of the distribution.
For better visualisation of the tail distribution including the interference effect,
the bin contents of all distributions were shifted by $+0.1$ before normalisation.}
\label{fig:CouplingComp}
\end{figure}

Figure~\ref{fig:CouplingComp} shows the generated mass distribution at particle level for a $Y$ quark with a mass of 900~\GeV, produced with a coupling strength of 0.5 and scaled to the LO cross-section. It is compared with the generated mass distributions reweighted to a coupling strength of 0.14 with and without the interference term, which is also scaled to the LO cross-section. For the case without interference, it was explicitly checked that events generated with one coupling
and reweighted to another target coupling result not only in the same VLQ mass distribution but also in the same
distributions of other kinematical variables when generated directly with this target coupling.
 
\section{Event selection and background estimation}
\label{sec:analysis}
This search focuses on final states with a leptonically decaying $W$ boson and a $b$-quark, originating from the decay of a
singly produced $Q$ quark. 
Events are required to have exactly one isolated identified lepton (electron or muon) with $\pt > 28$~\GeV\ that must be matched to the lepton selected by the trigger, large missing transverse momentum \mbox{$\met >$ 120~\GeV} from the escaping neutrino, and at least one central jet with $\pT > $ 25~\GeV\ satisfying the quality and kinematic criteria discussed in Section~\ref{sec:selection}. The requirement on the missing transverse momentum reduces the fraction of selected events originating from non-prompt or misidentified leptons as well as diboson events. In the following, unless stated otherwise, only events satisfying this selection, referred to as ``preselection'', are considered. If there are any forward jets in the event, their transverse momentum is required to be larger than 40~\GeV.
 
\subsection{Signal and control regions definition}
 
Events must have at least one $b$-tagged jet. The highest-$\pT$ jet in the event must be
$b$-tagged and have $\pt > 350~\GeV$. To further exploit the low multiplicity of high-$\pT$ jets in the signal process, an additional requirement is applied: events containing any jet with $\pT > 75~\GeV$ and $|\eta|<2.5$ and satisfying $\dR(\text{jet, leading jet}) < 1.2$
or $\dR(\text{jet, leading jet}) > 2.7$ are rejected.
This requirement reduces background from production of $\ttbar$ events,
which are characterised by a higher multiplicity of high-$\pT$ central jets than in signal events.
A requirement on the azimuthal separation between the lepton and the $b$-tagged leading jet, \mbox{|$\Delta \phi$ (lepton, leading jet)| $>$ 2.5}, as well as on the minimum distance $\dR$ between the lepton and any central jet, \mbox{$\dR$(lepton, jet) $>$ 2.0}, increases the signal-to-background ratio because, in signal signatures, leptons from the leptonic $W$-boson decays should be isolated and recoil against the $b$-quark jet in the event.
Furthermore, similar to $t$-channel single-top production, the single production of VLQs gives rise to a forward jet ($2.5 <|\eta|< 4.5$). Only events with at least one forward jet with \mbox{$\pt$ $>$ 40~\GeV} are considered. For a $Y$ signal with a mass between 800~\GeV\ and 2000~\GeV\ and a coupling strength of ${\sqrt{\cTwoL+\cTwoR} \approx 1/\sqrt{2}}$, the signal-to-background ratio ($S/B$) and the signal-to-background significance ratio ($S/\sqrt{B}$) in the SR are in the range 1.0--0.003 and 22.1--0.3 respectively. The acceptance times efficiency including the leptonic $W$ decay branching fractions\footnote{Events with leptonic $\tau$ decays are included.} for these $Y$ signals ranges from 0.7\% to 1.8\% in the SR.
 
The normalisation of $W$+jets and $\ttbar$ processes is partially constrained by fitting the predicted yields to data in CRs enriched in $W$+jets and $\ttbar$ events.
Two CRs are defined for this purpose, and also provide samples depleted in expected signal events. The selection requirements for the $W$+jets CR are the same as for the SR, except that each event is required to have exactly one $b$-tagged jet and the requirement on the azimuthal separation between the lepton and the $b$-tagged jet is reversed, \mbox{|$\Delta \phi$ (lepton, leading jet)| $\leq$ 2.5}. In addition, the $b$-tagged jet has a slightly lower transverse momentum requirement of \pT\ > 250~\GeV\ and
no hard or forward jet veto is applied.
The $W$+jets CR definition results in a composition of $W$+light-jets and $W$+heavy-flavour-jets final states similar to that in the SR.
The selection requirements for the $\ttbar$ CR are the same as for the SR, except that
the leading jet \pT\ must be greater than 200~\GeV\ and there must be at least one high-$\pT$ jet with
\pT\ > 75~\GeV\ and $|\eta|<2.5$ fulfilling either \mbox{$\dR$ (jet, leading jet) < 1.2} or
\mbox{$\dR$ (jet, leading jet) > 2.7}.
Table~\ref{tab:CRs} summarises the main selection criteria in the SR and the orthogonal CRs.
For $Y$/$T$ signals with masses of $\geq 800$~\GeV\ and a coupling strength of ${\sqrt{\cTwoL+\cTwoR} \approx 1/\sqrt{2}}$,
the contamination in the $\ttbar$ control region is at most 1\% and in the $W$+jets CR at most 0.6\%.
 
A mismodelling of the $W$+jets background is observed at high jet $p_{\mathrm{T}}$.
 
To correct for this mismodelling, the leading jet $p_{\mathrm{T}}$ distributions in data and MC-simulated $W$+jets events are compared after applying the preselection criteria and requiring that the leading jet is a $b$-tagged jet. The ratio of the distributions is taken as a scaling factor, which is applied to the simulated $W$+jets events in all kinematic distributions. The correction factors are between approximately 0.9 and 1.1 with statistical uncertainties of 4--10\% for a jet $p_{\mathrm{T}}$ below 500~\GeV, and 0.4--0.8 with a statistical uncertainty of about 11\% for higher \pT\ values. These
reweighting factors are treated as a systematic uncertainty in the final fit.\footnote{The residual  difference of about 10\% between the data and the SM simulation in the tail of the invariant mass distribution of the reconstructed VLQ candidate after applying the $W$+jets leading-jet $p_{\mathrm{T}}$ correction is included in this systematic uncertainty. }

\begin{table}[tbh]
\begin{center}
\caption{Summary of common preselection requirements and selection requirements for the SR compared to those for the \ttbar\ and $W$+jets CRs.
All other selection requirements are the same for all three regions.
}
\label{tab:CRs}
\vspace{2mm}
\begin{tabular}{l |ccc}
\toprule\toprule
\diagbox{Requirement}{Region}	 &SR  &\ttbar\ CR  &$W$+jets CR\\
\hline
\multicolumn{4}{c}{\textit{Preselection}}\\
\hline
Leptons &\multicolumn{3}{c}{1}\\
$\met$ &\multicolumn{3}{c}{$>$ 120~\GeV} \\
Central jets ($\pt$ $>$ 25~\GeV) &\multicolumn{3}{c}{$\geq$ 1}\\
\hline
\multicolumn{4}{c}{\textit{Selection}}\\
\hline
\\
$b$-tagged jets			&$\geq$ 1 &$\geq$ 1 & 1\\
 
Leading jet \pT\			& $>$ 350~\GeV\ &  $>$ 200~\GeV\ & $>$ 250~\GeV\ \\
Leading jet is $b$-tagged &Yes &Yes &Yes \\
$|\Delta\phi$(lepton, leading jet)|		& $>$ 2.5 & $>$ 2.5 & $\leq$ 2.5\\
 
Jets ($\pT > 75$~\GeV) with 		&&& \\
$\dR$ (jet, leading jet) $<$ 1.2 or &\multirow{2}{*}{0} &\multirow{2}{*}{ $\geq$ 1} &\multirow{2}{*}{--}\\
$\dR$ (jet, leading jet) $>$ 2.7		&&&\\
 
$\dR$ (lepton, jets)	& $>$ 2.0 & -- & $>$ 2.0	\\
 
Forward jets	($\pt$ $>$ 40~\GeV)			&$\geq$ 1 &$\geq$ 1 & --\\
\bottomrule\bottomrule
\end{tabular}
\end{center}
\end{table}

\subsection{Estimation of non-prompt and fake lepton backgrounds}
\label{sec:fake}
 
Multijet production results in hadrons, photons and non-prompt leptons that may satisfy the
lepton selection criteria and give rise to so called ``non-prompt and fake'' lepton backgrounds.
The multijet background normalisation and shape in the $m_\text{VLQ}$ distributions are estimated with a data-driven method, referred to as the Matrix Method~\cite{ATLAS-CONF-2014-058}.
This method uses the efficiencies of leptons selected using loose requirements (loose leptons) to pass the default tight lepton selection requirements.
The efficiencies are obtained in dedicated control regions enriched in real leptons or in non-prompt and fake leptons, and applied to events selected with either the loose or tight lepton definition to obtain the fraction of multijet events. \\
The fake-enriched control regions are defined using the preselection criteria, except that events with electrons are required to have a reconstructed transverse $W$ mass\footnote{The transverse $W$ mass \mtw\ is computed from the missing transverse momentum \vecMET and the charged lepton transverse momentum \ptll, and is defined as $\mtw=\sqrt{2\pTll E^{\mathrm{miss}}_{\mathrm{T}}(1-\cos\Delta\phi (\ptll,\vecMET))}$, where $\Delta\phi(\ptll,\vecMET )$ is the azimuthal angle between \ptll and \vecMET.} \mbox{$\mtw < 20~ \GeV$} and to have \mbox{$\met + \mtw < 60~ \GeV$}, and for events with muons it is required that the leading muon have $|d_{0} / \sigma(d_{0})| > 5$. The real lepton efficiencies are measured using the tag-and-probe method from $Z\rightarrow ee$ and $Z\rightarrow \mu\mu$ control regions. Further details can be found in Refs.~\cite{EXOT-2014-12, ATLAS-CONF-2014-058}.

\subsection{Signal candidate mass reconstruction}
 
In the SR, the invariant mass of the reconstructed VLQ candidate \mQ\ is used to discriminate the signal from the background processes. It is calculated from the leading $b$-tagged jet and
the decay products of the leptonically decaying $W$-boson candidate. The $W$-boson candidate is reconstructed by summing the four-momenta of the charged lepton and the neutrino.
To obtain the $z$-component of the neutrino momentum ($p_{z,\nu}$), the invariant mass of the lepton--neutrino system is set to the $W$-boson mass and the resulting quadratic equation is solved. If no real solution exists, the $\vec{E}_{\mathrm{T}}^{\mathrm{miss}}$ vector is varied by the minimum amount required to produce exactly one real solution.
If two real solutions are found, the one with the smaller $|p_{z,\nu}|$ is used. The $W$-boson candidate and the leading $b$-tagged jet are then used to reconstruct the $Q$ candidate. The mass resolutions for $Y$ signals with masses between
800~\GeV\ and 1600~\GeV\ for a coupling of ${\sqrt{\cTwoL+\cTwoR} \approx 1/\sqrt{2}}$ are 150--550~\GeV.
 
\begin{figure}[htb]
\centering
\subfigure[]{
\includegraphics[width=0.5\textwidth]{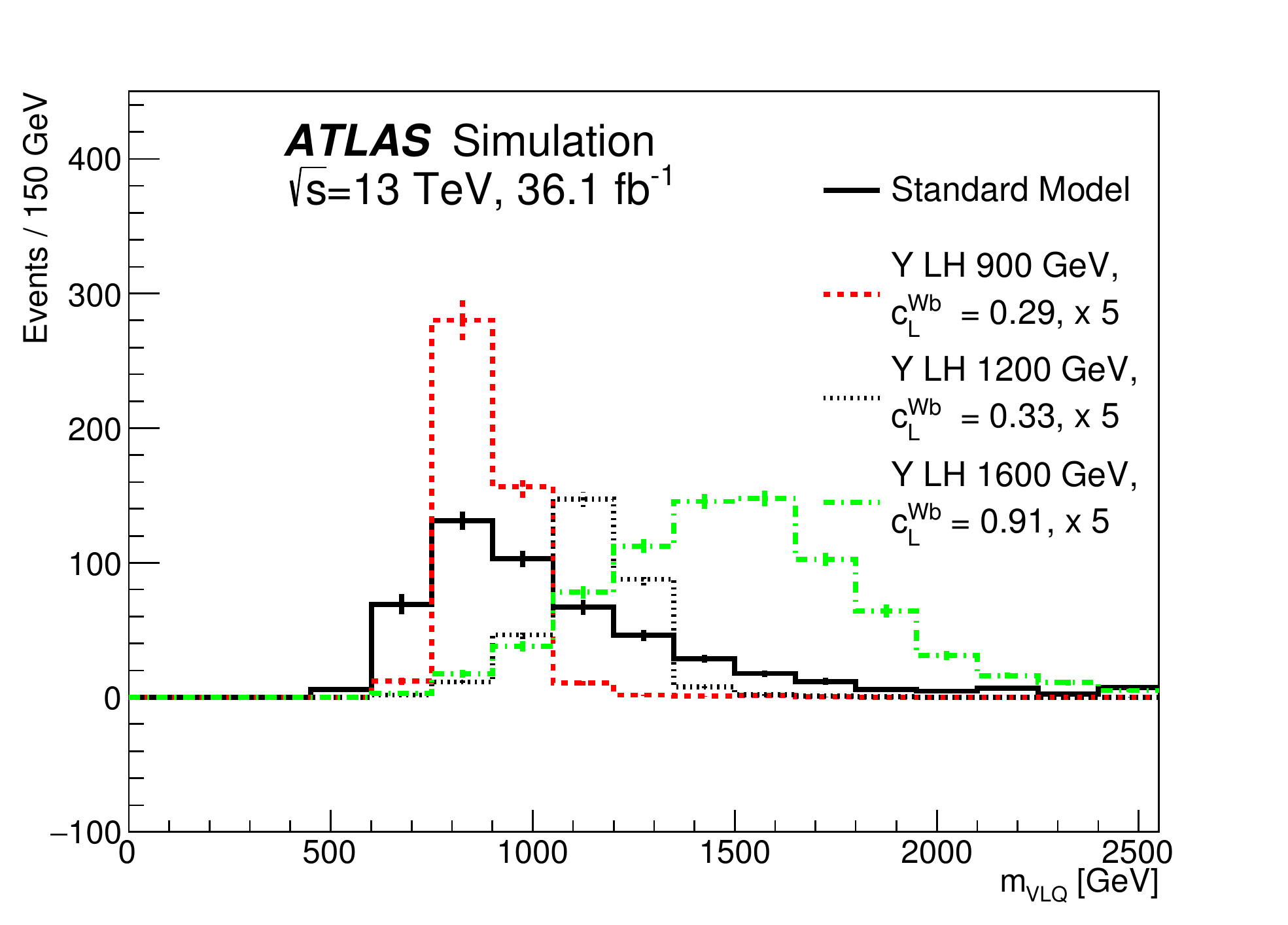}
\label{Fig:VLQ_AllSig_noint}
}
\hfill
\subfigure[]{
\includegraphics[width=0.5\textwidth]{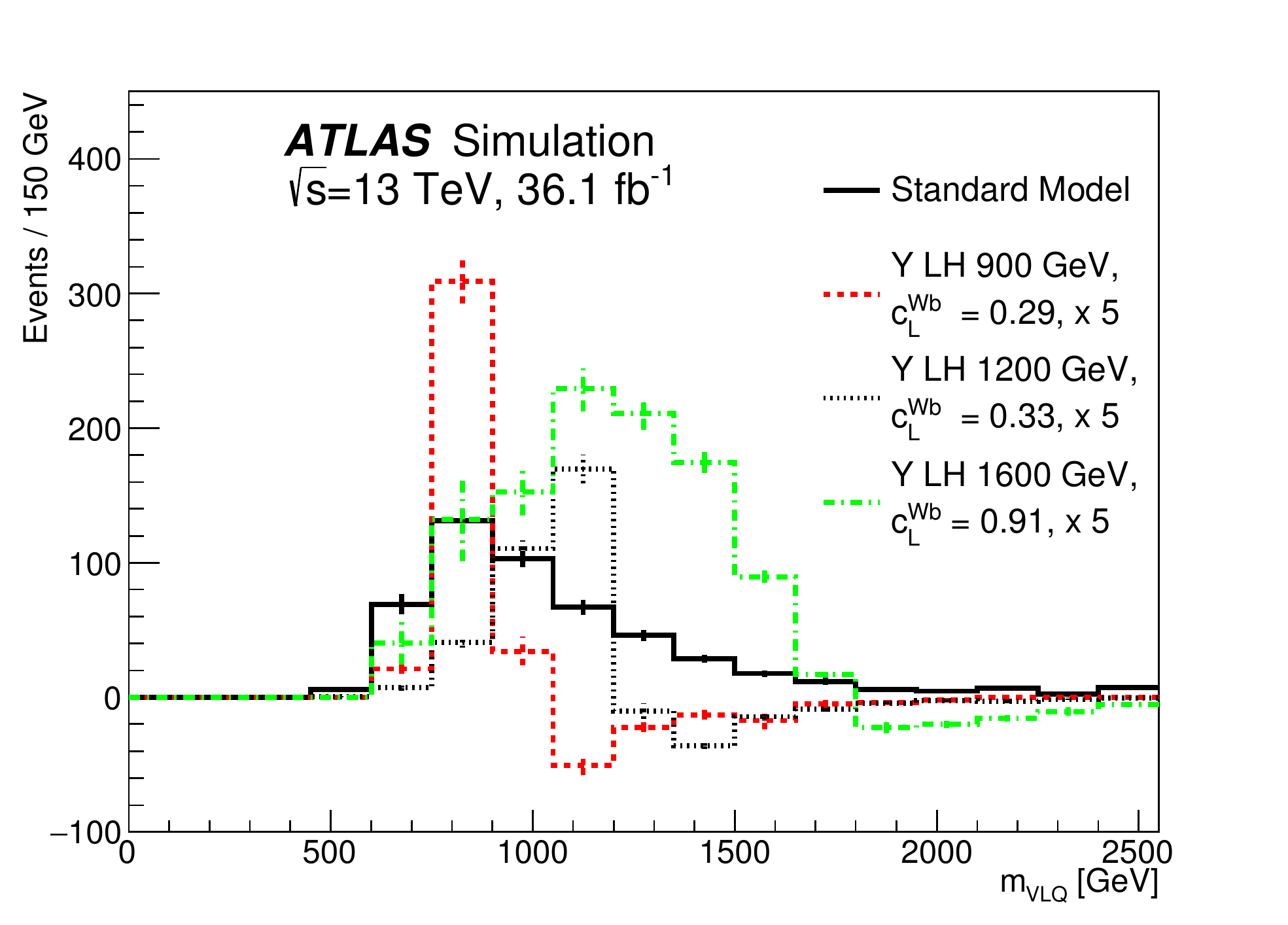}
\label{Fig:VLQ_AllSig_int}
}
\null
\caption{Distribution of VLQ candidate mass, $\mQ$, in the SR for three different signal masses \subref{Fig:VLQ_AllSig_noint} without
and \subref{Fig:VLQ_AllSig_int} with interference effects, for a left-handed $Y$ signal with a mass of 900~\GeV\ (dashed line), 1200~\GeV\ (dotted)
and 1600~\GeV\ (dash-dotted line) and a coupling of $\cL \approx 0.29$, $\approx 0.33$ and $\approx 0.91$ respectively, together with the total
SM background (solid line). The error bars represent the statistical uncertainties. The signal event yield is scaled by a factor of five.
Depending on the coupling and signal mass it is possible that negative entries occur in some bins of the signal-plus-interference $\mQ$
distribution due to the interference effect. The distributions for a right-handed and left-handed $Y$ signal without considering any
interference effects are the same.}
\label{fig:VLQ_AllSig}
\end{figure}

Figure~\ref{fig:VLQ_AllSig} shows the VLQ candidate invariant mass distribution in the SR for three
simulated left-handed $Y$ signal masses, 900~\GeV, 1200~\GeV\ and 1600~\GeV, with couplings of
${\cL \approx 0.29}$, $\approx 0.33$ and $\approx 0.91$ respectively,
without (left figure) and with (right figure) interference included,
together with the total SM background. The distribution provides good discrimination between signal
and background events in the SR. Depending on the coupling and signal mass it is possible that negative
entries occur in some bins of the signal-plus-interference $\mQ$ distribution due to the interference effect.
 
\section{Systematic uncertainties}
\label{sec:systematics}
Several sources of systematic uncertainty in this analysis can affect the normalisation of the signal and background and/or their corresponding \mQ\ distributions, which are used for the statistical study. They are included as nuisance parameters in the statistical analysis.
Sources of uncertainty include the modelling of the detector response, object reconstruction algorithms, uncertainty in the theoretical modelling of the signals and backgrounds, as well as the uncertainty arising from the limited size of the simulated event samples.
 
The following section describes each of the systematic uncertainties considered in the search.
Table~\ref{tab:SystSummary} presents a summary of all systematic uncertainties considered in the analysis. Leading sources of systematic uncertainty in the expected SM background are uncertainties that arise from the jet energy scale, flavour-tagging efficiencies ($b$, $c$ and light) as well as the background modelling, where $t\bar{t}$ generator uncertainties and single-top-quark DS/DR uncertainties are significantly constrained by the fit (see Section~\ref{sec:StatMethods}).

\begin{table}[ht!]
\centering
\caption{\label{tab:SystSummary} Systematic uncertainties considered in this analysis. An uncertainty that affects
normalisation only (cross-section only) for all processes and channels is denoted by ``N", whereas ``SN" means that the uncertainty affects both shape and normalisation and ``F" means a floating normalisation uncertainty.
Some of the systematic uncertainties are split into several components for a more accurate treatment. The relative systematic uncertainties in the inclusive expected SM background yields determined from the VLQ candidate invariant mass distribution after the fit to the background-only hypothesis are given in the last column in percentage.
The $t\bar{t}$ and $W$+jets background scaling-factor uncertainties (last two rows in the table) are the relative systematic uncertainties in the predicted $t\bar{t}$ and $W$+jets background respectively.
}
\small{
\begin{tabular}{llr}
\toprule\toprule
Systematic uncertainty & Type  &SM background [\%]\\
\midrule
Luminosity                  &  N &2.1 \\
Pile-up &SN &0.3 \\\midrule
\textit{Reconstructed objects: }                &  & \\
Electron efficiency, energy scale, resolution & SN &0.9 \\
Muon efficiency, momentum scale, resolution & SN &0.7 \\
Jet vertex tagger         & SN   & 0.1\\
Jet energy scale            & SN &6.4\\
Jet energy resolution       & SN &2.7\\ 
Missing transverse momentum  & SN &0.3\\
$b$-tagging efficiency for $b$-jets     & SN &0.8\\
$b$-tagging efficiency for $c$-jets     & SN &1.8\\
$b$-tagging efficiency for light-flavour jets     & SN &8.4\\ \midrule
\textit{Background model:}                 &   & \\
$t\bar{t}$ modelling: ISR/FSR & SN &0.2\\
$t\bar{t}$ modelling: generator & SN &3.8\\
$t\bar{t}$ modelling: parton shower/hadronisation & SN  &4.5\\
$t\bar{t}$ modelling: interfering background shape & S &0.3\\
Single-top cross-section    &  N &0.4\\
Single-top modelling: ISR/FSR & SN &0.04\\
Single-top modelling: generator & SN &0.3\\
Single-top modelling: DS/DR & SN &3.1\\
Single-top modelling: parton shower/hadronisation & SN  &1.6\\ 
$W$+jets modelling: generator & SN &0.8\\
$W$+jets modelling: reweighting &S &4.6 \\
$W$+jets heavy flavour &S &0.04\\
Diboson + $Z$+jets  normalisation  &  N &0.2\\
Multijet normalisation  &  N &3.8\\
Multijet reweighting &S &2.1\\
\midrule\midrule
$t\bar{t}$ background scaling factor & F &26\\
$W$+jets background scaling factor & F &19\\
\bottomrule\bottomrule
\end{tabular}
}
\end{table}

\subsection{Experimental uncertainties}
 
The uncertainty in the combined 2015+2016 integrated luminosity is $2.1\%$.
It is derived, following a methodology similar to that detailed in
Ref.~\cite{DAPR-2013-01}, and using the LUCID-2 detector for the baseline
luminosity measurements \cite{Avoni:2018iuv}, from calibration of the luminosity
scale using x-y beam-separation scans.
 
Experimental sources of systematic uncertainty arise from the reconstruction and measurement of jets~\cite{Aaboud:2017jcu}, leptons~\cite{ATLAS-CONF-2016-024, PERF-2015-10} and $\met$~\cite{PERF-2016-07}.
Uncertainties associated with leptons arise from the trigger, reconstruction, identification, and isolation efficiencies, as well as the lepton momentum scale and resolution, and are studied using $Z \rightarrow \ell^{+} \ell^{-}$ and $J/\psi \rightarrow \ell^{+} \ell^{-}$ decays in data. Uncertainties associated with
jets primarily arise from the jet energy scale, jet energy resolution, and the efficiency of
the jet vertex tagger requirement. The largest contribution is from the jet energy scale, where the dependence of the uncertainty on jet \pt and $\eta$, jet flavour, and pile-up is split into 21 uncorrelated components
that are treated independently in the analysis~\cite{Aaboud:2017jcu}. The systematic uncertainty in the \MET reconstruction is dominated by the uncertainties in the energy calibration and resolution of reconstructed jets and leptons, which are propagated to $\met$ and thus are included in the uncertainties in the corresponding objects.
In addition, uncertainties in the \pt\ scale and resolution of reconstructed tracks that are associated with the hard-scatter vertex but not matched to any reconstructed objects are included.
 
The efficiency of the flavour-tagging algorithm to correctly tag $b$-jets, or to mis-tag $c$-jets or light-flavour jets, is measured for each jet flavour. The efficiencies are measured in control samples of simulated events, and in data samples of  \ttbar\ events, $D^{*}$ mesons, and jets with impact parameters and secondary vertices consistent with a negative lifetime.
Correction factors are defined to correct the tagging rates in the simulation to match the efficiencies measured in the data control samples~\cite{PERF-2012-04, Aaboud:2018xwy}. The uncertainties associated with these measurements are factorised into statistically independent sources and include a total of six independent sources affecting $b$-jets and four independent sources affecting $c$-jets. Each of these uncertainties has a different dependence on jet \pt. Seventeen sources of uncertainty affecting light-flavour jets are considered, and depend on jet \pt and $\eta$. These correction factors are only determined up to a jet \pT\ of 300~\GeV\ for $b$- and $c$-jets, and \pt of 750~\GeV\ for light-flavour jets. Therefore, an additional uncertainty is included to extrapolate these corrections to jets with \pT\ beyond the kinematic reach of the data calibration samples used; it is taken to be correlated among the three jet flavours. This uncertainty is evaluated in the simulation by comparing the tagging efficiencies while varying, e.g., the fraction of tracks with shared hits in the silicon detectors or the fraction of fake tracks resulting from random combinations of hits, both of which typically increase at high jet \pT\
due to growing track multiplicity and density of hits within the jet. Finally, an uncertainty related to the
application of $c$-jet scale factors to $\tau$-jets is considered, but has a negligible impact in this analysis~\cite{Aaboud:2018xwy}.
 
The flavour-tagging systematic uncertainties are the leading sources of experimental uncertainties (added in quadrature, about 8.7\% in the expected background yield in the SR).
Other large detector-specific uncertainties arise from jet energy scale uncertainties (about a 6.4\% effect on the expected background yield) and jet energy resolution uncertainties (2.7\% in the expected background yield).
The total systematic uncertainty associated with \MET\ reconstruction is about 0.3\% in the SR. The combined effect of all these uncertainties results in an overall normalisation uncertainty in the SM background of approximately 6.3\% taking into account correlations between the different systematic uncertainties.
 
For the data-driven multijet background, which has a very small contribution in the SR and CRs, a 100\% normalisation uncertainty is used, to fully cover discrepancies between the observed data and the SM expectation
in multijet-background-enriched regions. The large statistical uncertainties associated with the multijet background prediction, which are uncorrelated bin-to-bin in the final discriminating variable, do not cover shape differences in the multijet background electron \pT\ distribution. This mismodelling is corrected by determining reweighting factors in a multijet-background-enriched region which are used as additional shape uncertainties in the final discriminant. These reweighting factors are obtained for electrons with $|\eta| < 1.2$ and $|\eta| > 1.2$ separately in a region requiring the same selection requirements as the preselection, but loosening the minimum $\met$ requirement to 20~\GeV\ and requiring the leading jet is a $b$-jet.

\subsection{Theoretical modelling uncertainties}
A number of systematic uncertainties affecting the modelling of $t\bar{t}$ and single-top-quark processes as described in Section~\ref{sec:bkgmodelling} are considered: uncertainties associated with the modelling of the
ISR and FSR, uncertainties associated with the choice of NLO generator, modelling uncertainties in single-top-quark production (for $t$-channel) based on comparison of the nominal sample with an alternative MC sample described in Section~\ref{sec:bkgmodelling}, differences between single-top-quark $Wt$ samples produced using the diagram subtraction scheme and $Wt$ samples produced using the diagram removal scheme, as well as an uncertainty due to the choice of parton shower and hadronisation model.
The $\ttbar$ background normalisation is a free parameter in the fit, while the normalisation of the single-top background has an uncertainty of 6.8\%~\cite{Kidonakis:2010ux}.
 
Uncertainties affecting the modelling of the $Z$+jets background and diboson background processes include a 5\% effect from their respective normalisations to the theoretical NNLO cross-sections~\cite{melnikov:2006kv, campbell:1999ah, Anastasiou:2003ds}.
Since both these backgrounds are very small, this uncertainty is applied to the sum of the predicted $Z$+jets and
diboson background processes.
The $W$+jets background normalisation is a free parameter in the fit.
The $W$+light-jets and $W$+heavy-flavour-jets predictions have similar \mQ\ distributions in the SR and CRs. Since the predicted
ratios of $W$+light-jets to $W$+heavy-flavour-jets events in the SR and CRs are similar, but not identical, a systematic uncertainty is derived by comparing the shape of the complete $W$+jets sample with the $W$+heavy-flavour-jets portion alone.
In addition, alternative $W$+jets samples were generated using {\scshape Madgraph}+{\Pythia 8} and compared after applying the    preselection criteria plus requiring that the leading jet is a $b$-tagged jet.
 
To account for the mismodelling of the leading-jet $p_{\mathrm{T}}$ spectrum in $W$+jets events, reweighting factors are obtained at preselection for $W$+jet events. The \mQ\ distributions with and without these $W$+jets jet-$\pT$ correction factors applied to $W$+jet events are compared in the SR and CRs and used to quantify the systematic uncertainty in the \mQ\ shape of $W$+jets events in the fit.
 
All normalisation uncertainties in the different background processes are treated as uncorrelated. For background estimates based on simulations, the largest sources of theoretical modelling uncertainties are due to the choice of parton shower and hadronisation model (2--4\%), the choice of generator (about 1--3\% in the expected background yield) and varying the parameters controlling the initial- and final-state radiation (about 0.1\% in the expected background yield), where the theoretical modelling uncertainties from $\ttbar$ contribute the most.
 
The systematic uncertainties in the modelling of the high-mass $Y$/$T$ signal sample which
correspond to the choice of PDF set are evaluated following the PDF4LHC15 prescription~\cite{Butterworth:2015oua}.
No further systematic uncertainties in the signal modelling and no uncertainties in the NLO signal production cross-section are considered.
In addition, a systematic uncertainty of about 2.5\% is applied to cover small differences in the reconstructed VLQ mass between signal samples passed through the full simulation of the detector and signal samples produced with the faster simulation (see Section~\ref{sec:modelling}).
 
The ATLAS MC production used in this analysis does not contain simulated events
from the SM contributions that lead to interference with the VLQ signal.
Therefore, these SM contributions can not be explicitly considered in the
background modelling of the fit. A recent MC production at
reconstruction level using the four-flavour scheme for one mass point for a left-handed $Y$ quark
shows that the \mQ\ distribution of the interfering SM contribution is similar but not identical
to that of the other background contributions ($W$+jets, $\ttbar$, single top).
To account for the presence of interfering SM contributions in the fit,
an additional shape uncertainty is applied to the $\ttbar$ \mQ\ template,
which leads to an uncertainty of $0.2\%$ in the $\ttbar$ yield.

 
\section{Results}
\label{sec:result}

\subsection{Statistical interpretation}
\label{sec:StatMethods}
 
A binned maximum-likelihood fit to the data is performed to test for the presence of a signal.
A separate fit is performed for each signal hypothesis with given mass and couplings.
The inputs to the fit are the distributions of reconstructed VLQ candidate mass $\mQ$ in the SR and the two CRs.
The binned likelihood function ${\cal L}(\mu,\theta)$ is constructed as a product of Poisson probability
terms over all  $\mQ$ bins considered in the search.
It depends on the signal-strength parameter $\mu$, a multiplicative factor to the theoretical signal production
cross-section, and $\theta$, a set of nuisance parameters that encode the effect of systematic uncertainties in
the signal and background expectations and are implemented in the likelihood function as Gaussian constraints, as well as on the two scale factors for the free-floating \ttbar and $W$+jets SM background normalisations. Uncertainties in each bin of the $\mQ$ distributions due to the finite numbers of events in the simulation samples
are included using dedicated fit parameters and are propagated to $\mu$.
The nuisance parameters $\theta$ allow variations of the expectations for signal and background according to
the corresponding systematic uncertainties, and their fitted values $\hat{\theta}$ correspond to the deviations
from the nominal expectations which globally provide the best fit to the data.
This procedure reduces the impact of systematic uncertainties on the search sensitivity by taking
advantage of the well-populated background-dominated CRs included in the likelihood fit.
It also allows the CRs to improve the description of the data.
 
The test statistic $q_\mu$ is defined as the profile log-likelihood ratio:
$q_\mu = -2\ln({\cal L}(\mu,\hat{\hat{\theta}}_\mu)/{\cal L}(\hat{\mu},\hat{\theta}))$,
where $\hat{\mu}$ and $\hat{\theta}$ are the values of the parameters that
maximise the likelihood function (with the constraint $0\leq \hat{\mu} \leq \mu$),
and $\hat{\hat{\theta}}_\mu$ are the values of the nuisance parameters that maximise
the likelihood function for a given value of $\mu$.
In the absence of any significant deviation from the background expectation,
$q_\mu$ is used in the CL$_\text{s}$ method~\cite{Junk:1999kv,Read:2002hq} to set an upper limit
on the signal production cross-section times branching ratio at the 95\% CL. For a given signal scenario,
values of the production cross-section (parameterised by $\mu$) yielding CL$_\text{s} < 0.05$,
where CL$_\text{s}$ is computed using the asymptotic approximation~\cite{Cowan:2010js},
are excluded at 95\% CL.
 
\subsection{Fit results}
 
\begin{figure}[htb]
\begin{center}
\includegraphics[width=0.6\textwidth]{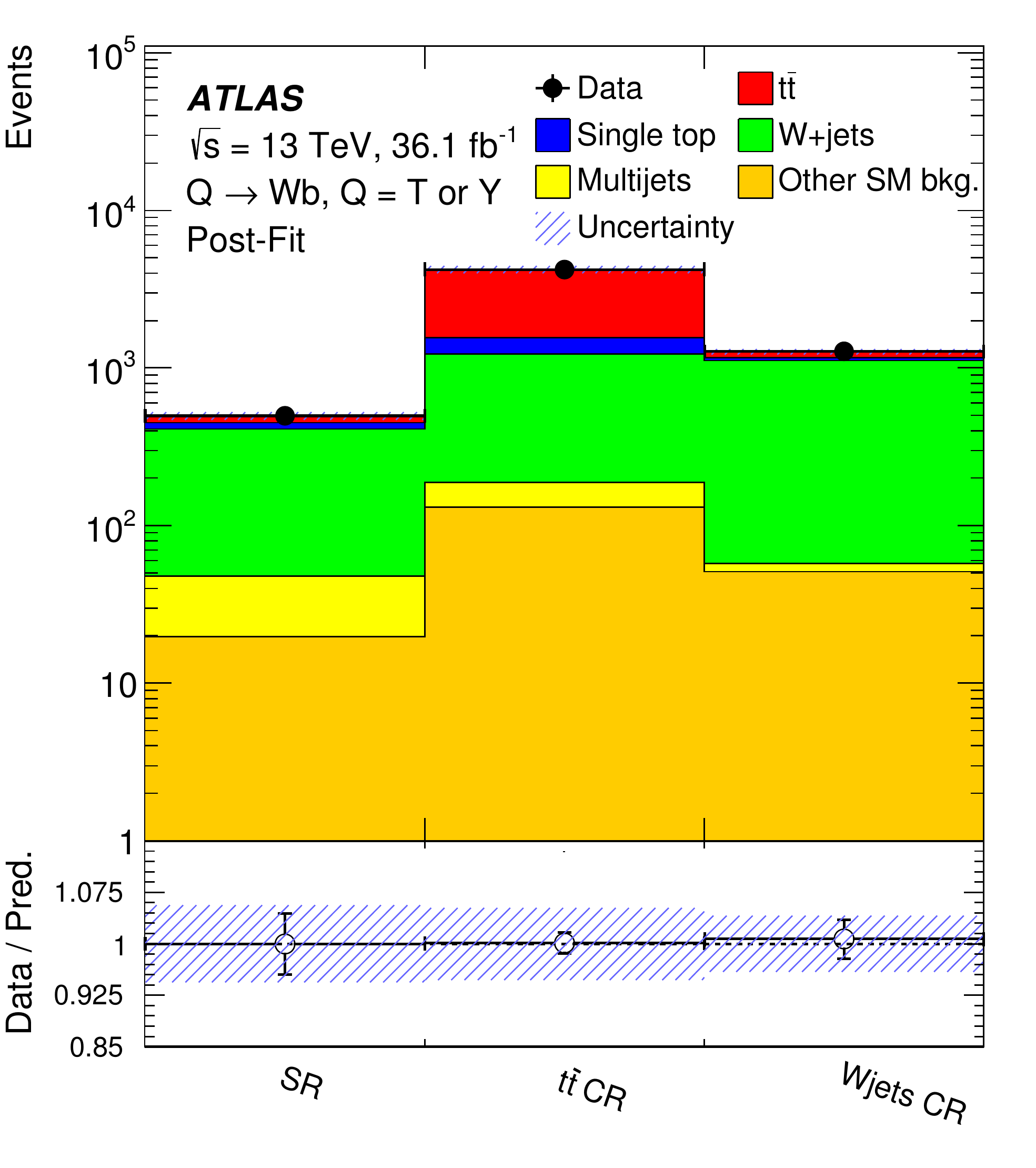}
\caption{Observed background yields in the SR
and in the two CRs after the fit to the data in the control regions and the signal region under the background-only hypothesis.
The lower panel shows the ratio of data to the fitted background yields. The error bars, being smaller than the size of the data points
and hence not visible in the top part of the plot, represent the statistical uncertainty in the data. The band represents the total
(statistical and systematic) uncertainty after the maximum-likelihood fit.
\label{fig:Summary_postFit}}
\end{center}
\end{figure}

\begin{figure}[htb]
\centering
\subfigure[]{
\includegraphics[width=0.45\textwidth]{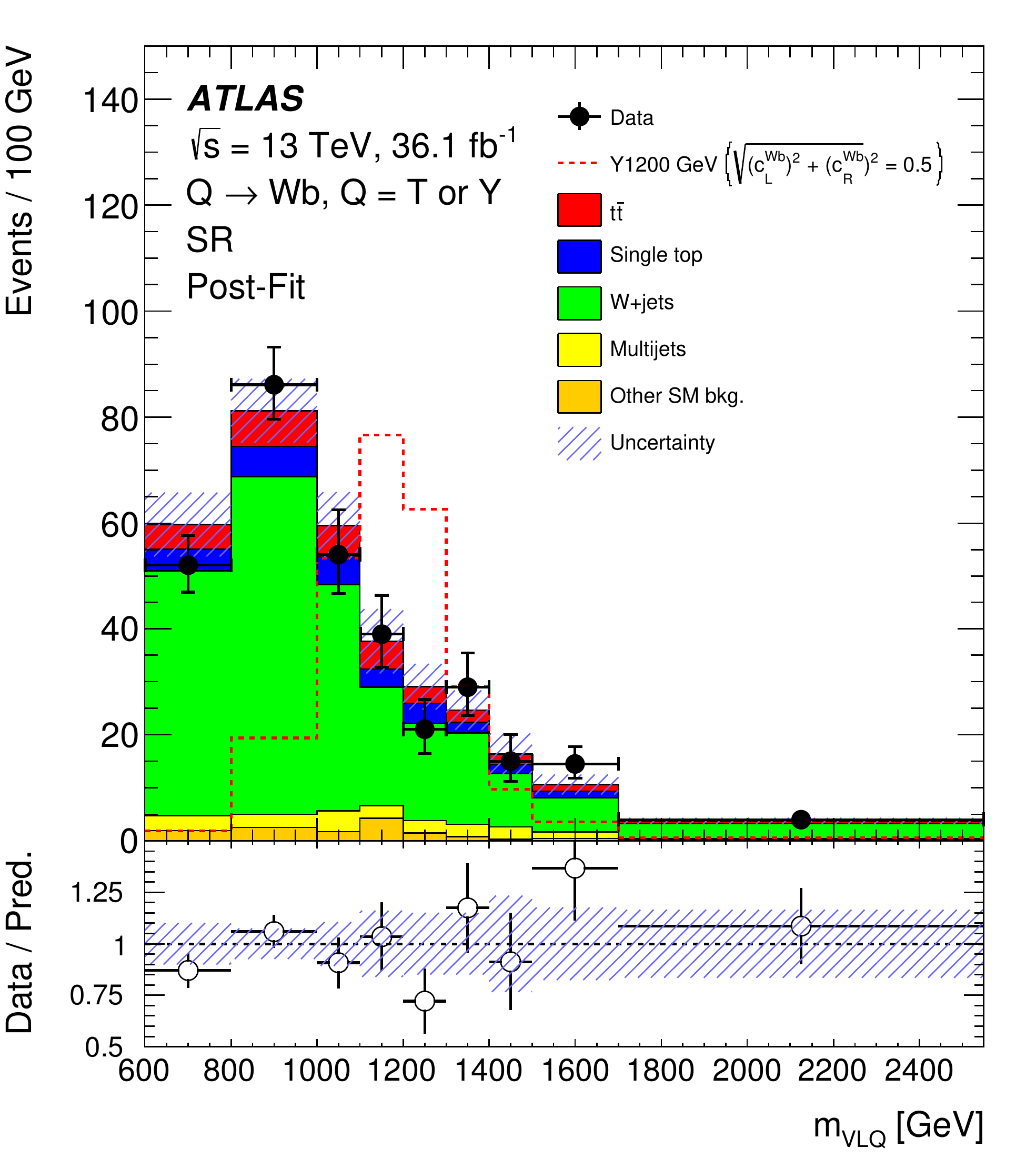}
\label{Fig:SR_postFit}
}
\hfill
\subfigure[]{
\includegraphics[width=0.45\textwidth]{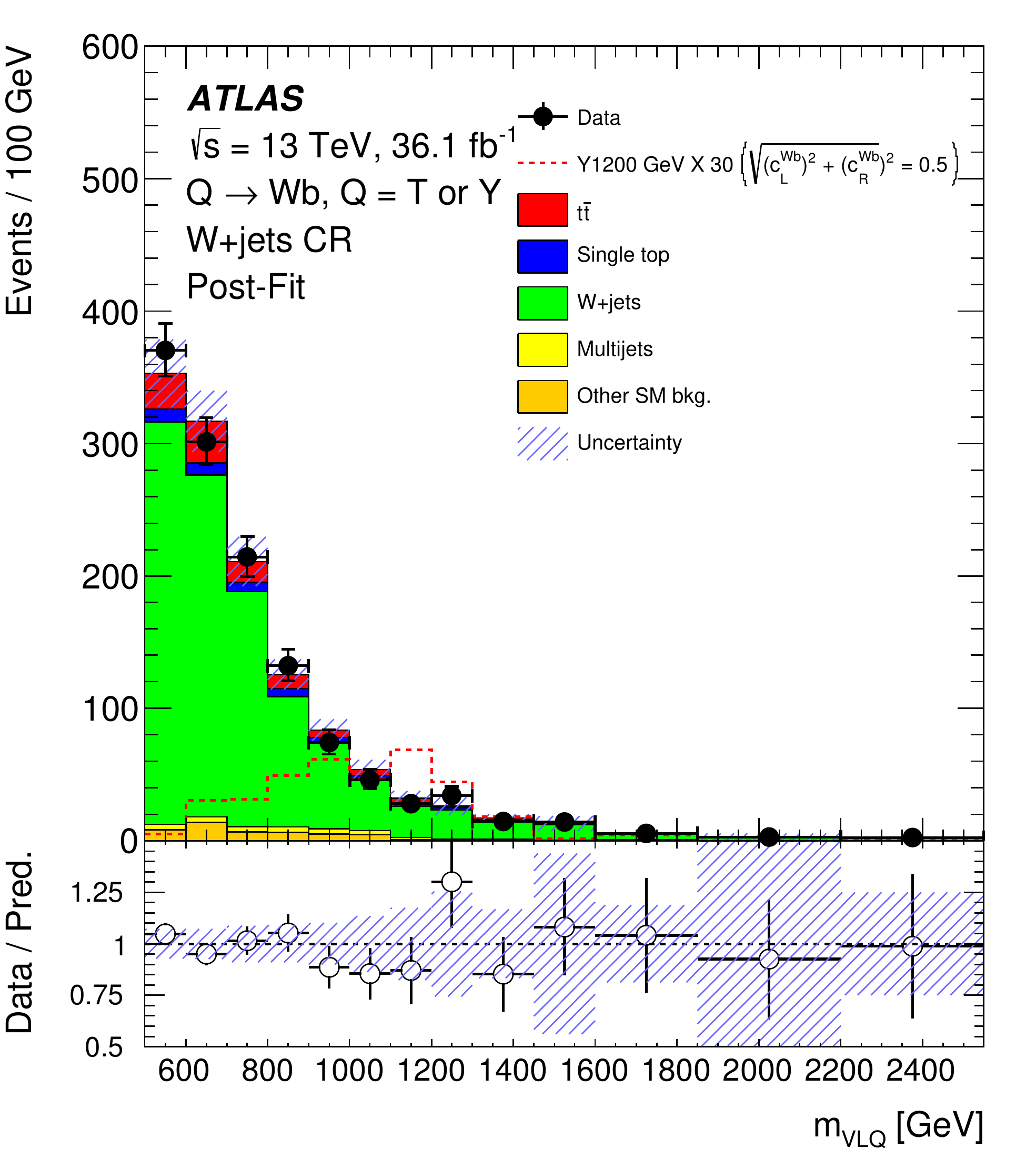}
\label{Fig:WjetsCR_postFit}
}
\hfill
\subfigure[]{
\includegraphics[width=0.45\textwidth]{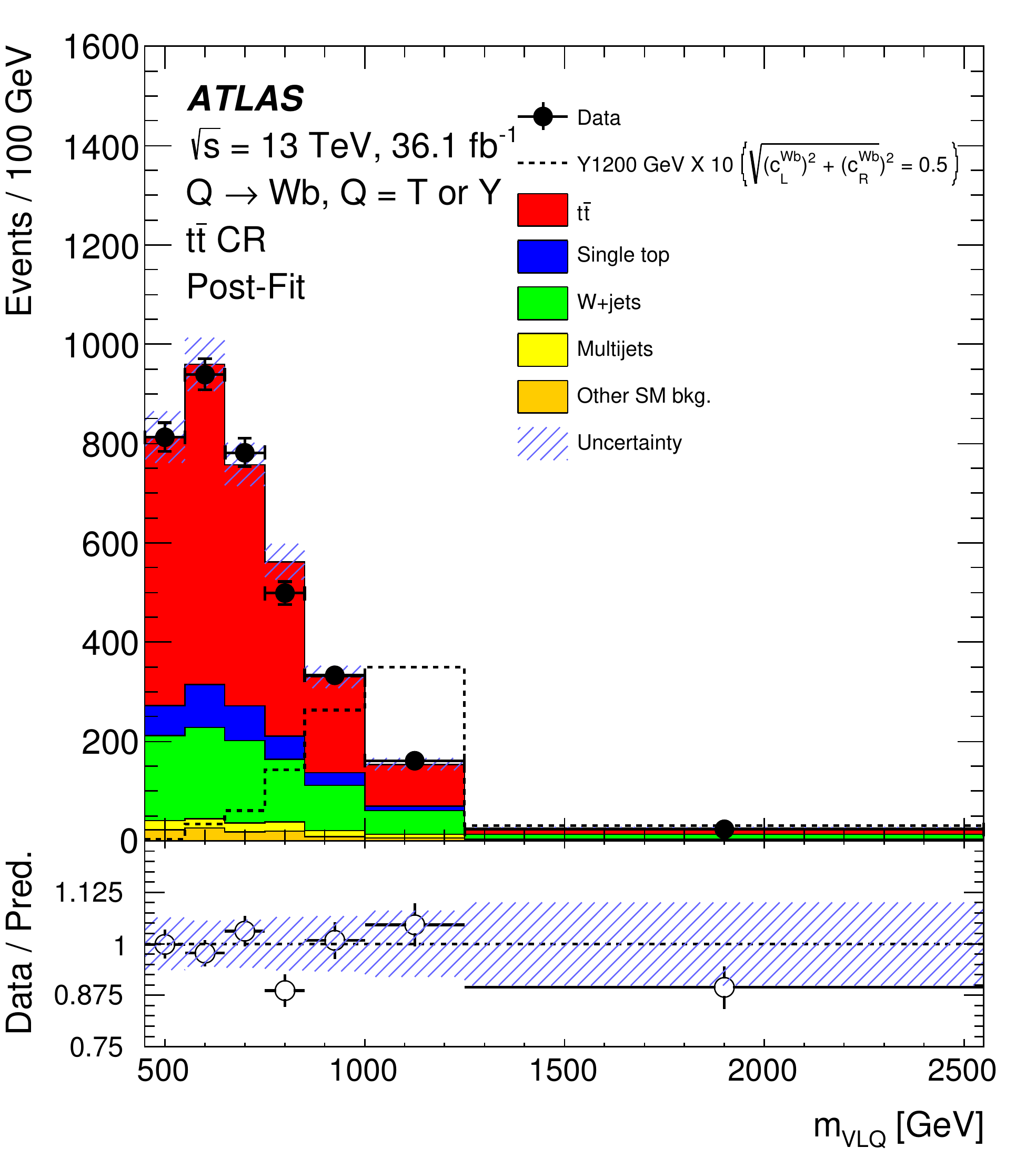}
\label{Fig:ttbarCR_postFit}
}
\null
\caption{Distribution of the VLQ candidate mass, $\mQ$, in \subref{Fig:SR_postFit} the SR, \subref{Fig:WjetsCR_postFit} the $W$+jets CR, and \subref{Fig:ttbarCR_postFit} the \ttbar CR, after the fit to the background-only hypothesis. The first and last bin include the underflow
and overflow respectively.
The lower panels show the ratios of data to the fitted background yields. The error bars represent
the statistical uncertainty in the data. The band represents the total systematic uncertainty after
the maximum-likelihood fit. An example distribution for a $Y$ signal with a coupling of
$\sqrt{\cTwoL+\cTwoR} \approx 0.5$
without considering any interference effects is overlaid; for better visibility, it is multiplied by
a factor of 30 in the $W$+jets CR and by a factor of 10 in the \ttbar CR.
While the total uncertainty decreases when performing the fit, the
total uncertainty in the bins around 1450-1600 GeV and 1850-2200 GeV
in \subref{Fig:WjetsCR_postFit} does not decrease due to significant
statistical MC uncertainties in these two bins.
}
\label{Fig:fitresults_postfit}
\end{figure}

The background-only fit results for the yields in the SR and the two CRs are shown in Figure~\ref{fig:Summary_postFit}.
Figure~\ref{Fig:fitresults_postfit} presents the \mQ\ distributions after the background-only fit
in the SR and the two CRs with the SR binning as used in the background-only fit. The overall $t\bar{t}$ ($W$+jets)
normalisation is adjusted by a factor of 0.95 $\pm$ 0.26 (1.18 $\pm$ 0.19), where 0.26 (0.19) is the total uncertainty
in the normalisation. An example distribution for a right-handed $Y$ signal and a coupling of
\mbox{$\cR \approx 0.5$}
is overlaid, which illustrates what such a signal would look like. Good agreement between the data and the SM backgrounds is found,
in particular in the SR for the \mQ\ distribution, where no peak above the expected SM background is observed.
 
The numbers of data events in the SR and CRs, and the event yields after fitting the background-only hypothesis to
data, together with their systematic uncertainties, are listed in Table~\ref{Tab::evnt_yields}.

\begin{table}[htb]
\begin{center}
\caption{\label{Tab::evnt_yields}
Event yields in the SR and the \ttbar and $W$+jets CRs after the fit to the background-only hypothesis.
The uncertainties include statistical and systematic uncertainties.
Due to correlations among the SM backgrounds and the corresponding nuisance parameters, the uncertainties in the individual background components can be larger than the uncertainty in the sum of the background, which is strongly constrained by the data.}
\sisetup{round-mode = places, round-precision=1, retain-explicit-plus=true, group-digits=integer, group-minimum-digits=5}
\begin{tabular}{ c
S[table-format=3.1, table-number-alignment=right] @{\,}@{$\pm$}@{\,} S[table-format=2.1, table-alignment=right]
S[table-format=4.1, table-number-alignment=right] @{\,}@{$\pm$}@{\,} S[table-format=3.2, table-alignment=right]
S[table-format=4.1, table-number-alignment=right] @{\,}@{$\pm$}@{\,} S[table-format=2.1, table-alignment=right]
}
\toprule\toprule
Source	   & \multicolumn{2}{c}{SR} & \multicolumn{2}{c}{\ttbar CR} & \multicolumn{2}{c}{$W$+jets CR} \\
\midrule
\ttbar & 58 & 21 & 2715 &295 &100 &29\\
Single top   & 29 &15 &271 &118 &34 &18 \\
$W$+jets  &373  &45 &1052 &143 &1077 &84 \\
Multijet $e$ &22 &20 &35 &40 &0 &4 \\
Multijet $\mu$ &7 &7 &92 &71 &26 &20 \\
$Z$+jets, diboson &20 &5 &102 &20 &50 &8 \\
\ttbar $V$ &0.3 &0.1 &21 &3 &1.6 &0.3 \\
\ttbar $H$ &0 &0 &7 &1 &0.2 &0.1 \\
Total & 500 &30 &4300 &210 &1290 &70 \\
\midrule
Data   & \multicolumn{2}{l}{497} & \multicolumn{2}{l}{4227} & \multicolumn{2}{l}{1274} \\
\bottomrule\bottomrule
\end{tabular}
\end{center}
\end{table}

\clearpage
\newpage
\subsection{Limits on the VLQ production}
\label{limits}
When allowing for the signal presence, no significant deviation from the expected SM background is found.
In all models considered in this search ($T$ singlet model, right-handed $Y$
in a $(B,Y)$ doublet model, left-handed $Y$ in a $(T,B,Y)$ triplet model),
interference effects with SM contributions affect the $\mQ$ distribution (see Section~\ref{sec:Signal}).
The effects of the interfering SM contributions ($\sigma_{\textrm{SM}}$, see Eq.~(\ref{eq:xsec})) in the fit are treated as a systematic
uncertainty in the background modelling (see Section~\ref{sec:systematics}).
Therefore, only the interference effect itself ($\sigma_{\textrm{I}}$) is
explicitly taken into account in the signal template.
For the left-handed $Y$ and the $T$-singlet case, the size and $\mQ$
distribution of the interfering SM contributions are estimated in three ways:
\begin{enumerate}
\item Using the shape of the reweighted template ($\sigma_{\textrm{VLQ}} + \sigma_{\textrm{I}}$).
\item Using simulated events in the four- and five-flavour schemes at particle level,
with the SR requirements applied.
\item Using the fully-reconstructed MC simulated events mentioned in Section~\ref{sec:systematics}.
\end{enumerate}
In the four-flavour scheme, the yield and the $\mQ$ distribution both agree within statistical uncertainties for the
left-handed $Y$ in a $(T,B,Y)$ triplet model and the $T$ singlet model. For the left-handed $Y$, the yields
in the four- and five-flavour schemes differ by a factor of about two, while
the $\mQ$ distributions in both schemes are very similar.
A background-only fit in the SR and CRs shows that the interfering SM
contribution, the shape of which is taken from the fully reconstructed
MC simulation mentioned above, is in agreement with the size used to
simulate the interference templates ($\sigma_{\textrm{I}}$) and can affect the total
postfit background yield by about $4~\%$.
This effect can be accounted for by adding the shape of the
interfering SM background as an additional systematic uncertainty in
the $\ttbar$ template (see Section ~\ref{sec:systematics}). Studies show that the expected
and observed limits change by significantly less than one standard deviation
with the addition of this systematic uncertainty.
For the right-handed $Y$ in a $(B,Y)$ doublet model, the interfering SM
background contributions are much smaller than other background
contributions in the SR and the CRs and are therefore
negligible. Nonetheless, the non-simulated SM contributions mentionned above, which
would lead to interference with a left-handed $Y$ in a $(T,B,Y)$ triplet model
or a $T$ singlet quark, are non-negligible and are therefore taken into account
in the fit by the same additional systematic uncertainty in the $\ttbar$ template.
Since the interfering SM contributions are not explicitly taken into
account in the fit, upper limits on the total cross-section for
$p p \rightarrow Wbq$,
$\sigma_{\textrm{tot}}=\sigma_{\textrm{VLQ}} + \sigma_{\textrm{I}} + \sigma_{\textrm{SM}}$,
times branching ratio can not be determined, but limits on the coupling value of the
vector-like $T$ or $Y$ quark to $Wb$ in a given model based on
$\sigma_{\textrm{VLQ}} + \sigma_{\textrm{I}}$ are set.
 
To set a coupling-value limit, the following iterative procedure is performed:
for a fixed $Q$ mass hypothesis and for a given coupling value $c^{Wb}$,
a $\mQ$ signal-plus-interference template $h_{\text {VLQ+I}}(\mQ;c^{Wb})$
containing the VLQ ($\sigma_{\text {VLQ}}$) and the interference contribution
($\sigma_{\text {I}}$) (but not the interfering SM contribution ($\sigma_{\textrm{SM}}$))
is constructed by reweighting the default VLQ-only signal template
$h_{\text {VLQ}}(\mQ;c^{Wb}_{\mathrm{def}})$ for a default coupling value ($c^{Wb}_{\mathrm{def}}$ = $c_0$) using the ratio $r$ (see Eq.~(\ref{eq:ratio})) defined in Section~\ref{sec:Signal}.
The maximum-likelihood fit to signal plus background is performed with the signal template $h_{\text {VLQ+I}}(\mQ;c^{Wb})$,
and an upper limit on $\sigma_{\text {VLQ}}+\sigma_{\text {I}}$ is determined.
The $T$-quark branching ratio is set to $\mathcal{B}$$(T \rightarrow Wb) = 0.5$,\footnote{For the $T$ singlet model, $\mathcal{B}(T \rightarrow Wb)= 0.5$ is a very good approximation in the mass and coupling ranges relevant to this search.}
whereas $\mathcal{B}$$(Y \rightarrow Wb)= 1$ is used for the $Y$ quark.
The theoretical cross-section $\sigma_{\text {VLQ}}$ is taken from Ref.~\cite{Matsedonskyi:2014mna},
where the NLO $Wb$ fusion cross-section is calculated in the NWA.
With rising $Q$ mass and coupling value $c^{Wb}$, the $Q$ width becomes sizeable and the NWA calculation
is no longer a good approximation. Therefore, the following correction factor applies to the
theoretical cross-section prediction:
\begin{linenomath*}\begin{equation*}
\frac{\sigma_{\mathrm{LO,noNWA}}}{\sigma_{\mathrm{LO,NWA}}}= C_{\mathrm{NWA}},
\end{equation*}\end{linenomath*}
where $\sigma_{\mathrm{LO,noNWA}}$ is the LO cross-section without the NWA and
$\sigma_{\mathrm{LO,NWA}}$ the LO cross-section with the NWA, both calculated with the
{\scshape Madgraph5}\_a{\scshape MC@NLO} 2.2.3~\cite{Alwall:2014hca} generator. It is assumed that
$C_{\mathrm{NWA}}$ is the same to a good approximation for the calculation of the NLO
cross-section. These correction factors reduce the predicted $\sigma_{\text {VLQ}}$ value.
The reduction becomes stronger with increasing mass and coupling value and is about $40\%$
at a $Q$ mass of 1500~\GeV\ and a coupling value of 0.9.
From the upper limit on $\sigma_{\text {VLQ}}+\sigma_{\text {I}}$, a corresponding coupling value
${c^{Wb}}'$ is calculated, a new signal template $h_{\text{VLQ+I}}(\mQ;{c^{Wb}}')$ is constructed using the
reweighting technique described above, and the fit repeated until convergence is observed in the
coupling value ${c^{Wb}}'$. It is explicitly checked that the result of the iterative procedure
does not depend on the choice of starting value for $c^{Wb}$, by repeating the full iterative process with a lower or higher starting value than the one at convergence. If the coupling converges to a value smaller than the signal-production value of 0.5, the iterative procedure is repeated with a coupling much lower than the value at convergence. A systematic uncertainty of about 2.5\% for the coupling reweighting and a shape uncertainty for the interference contribution are assigned to this procedure.
 
\begin{figure}[htb]
\centering
\subfigure[]{
\includegraphics[width=0.45\textwidth]{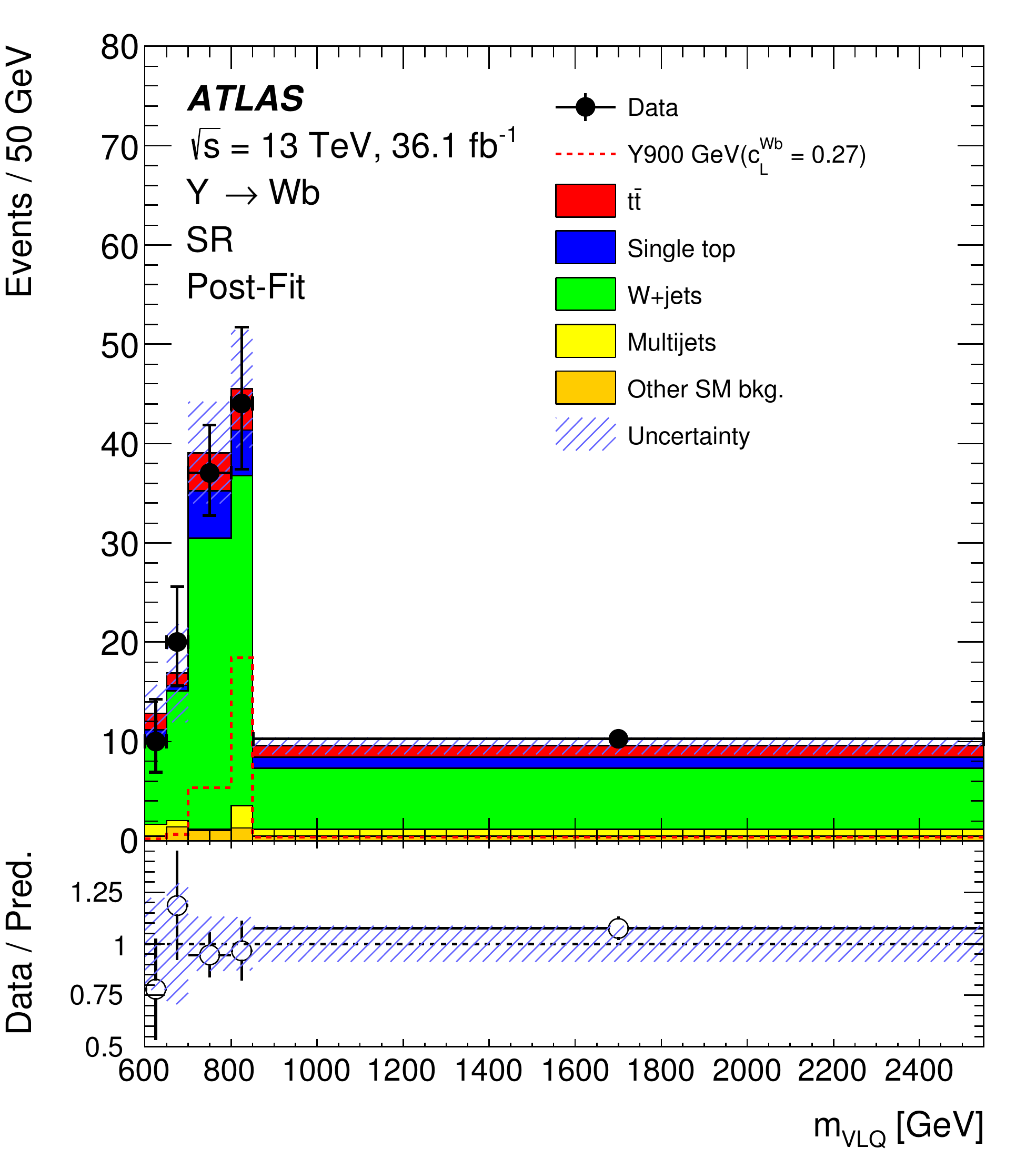}
\label{Fig:VLQ_Y1}
}
\hfill
\subfigure[]{
\includegraphics[width=0.45\textwidth]{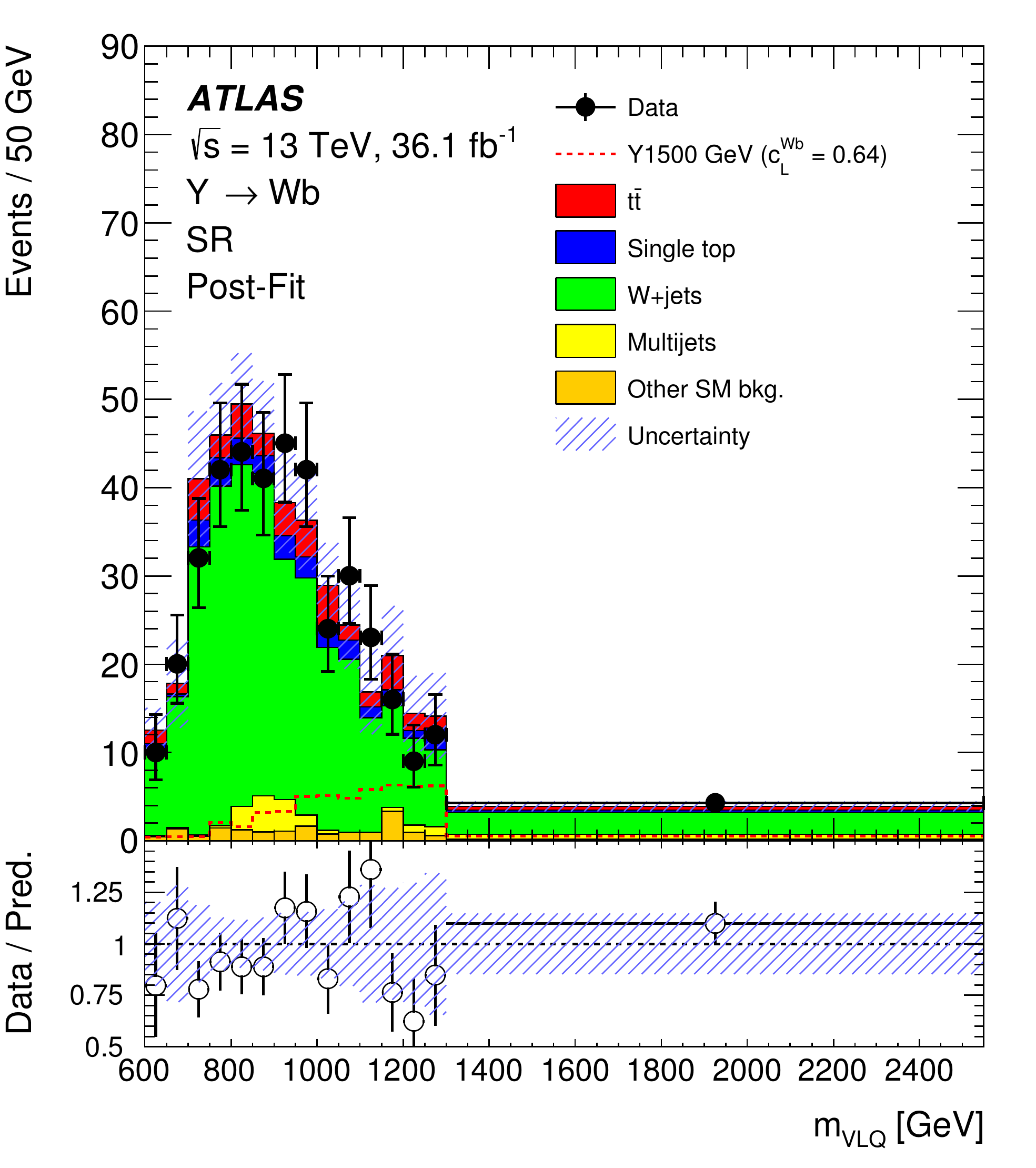}
\label{Fig:VLQ_Y2}
}
\hfill
\subfigure[]{
\includegraphics[width=0.45\textwidth]{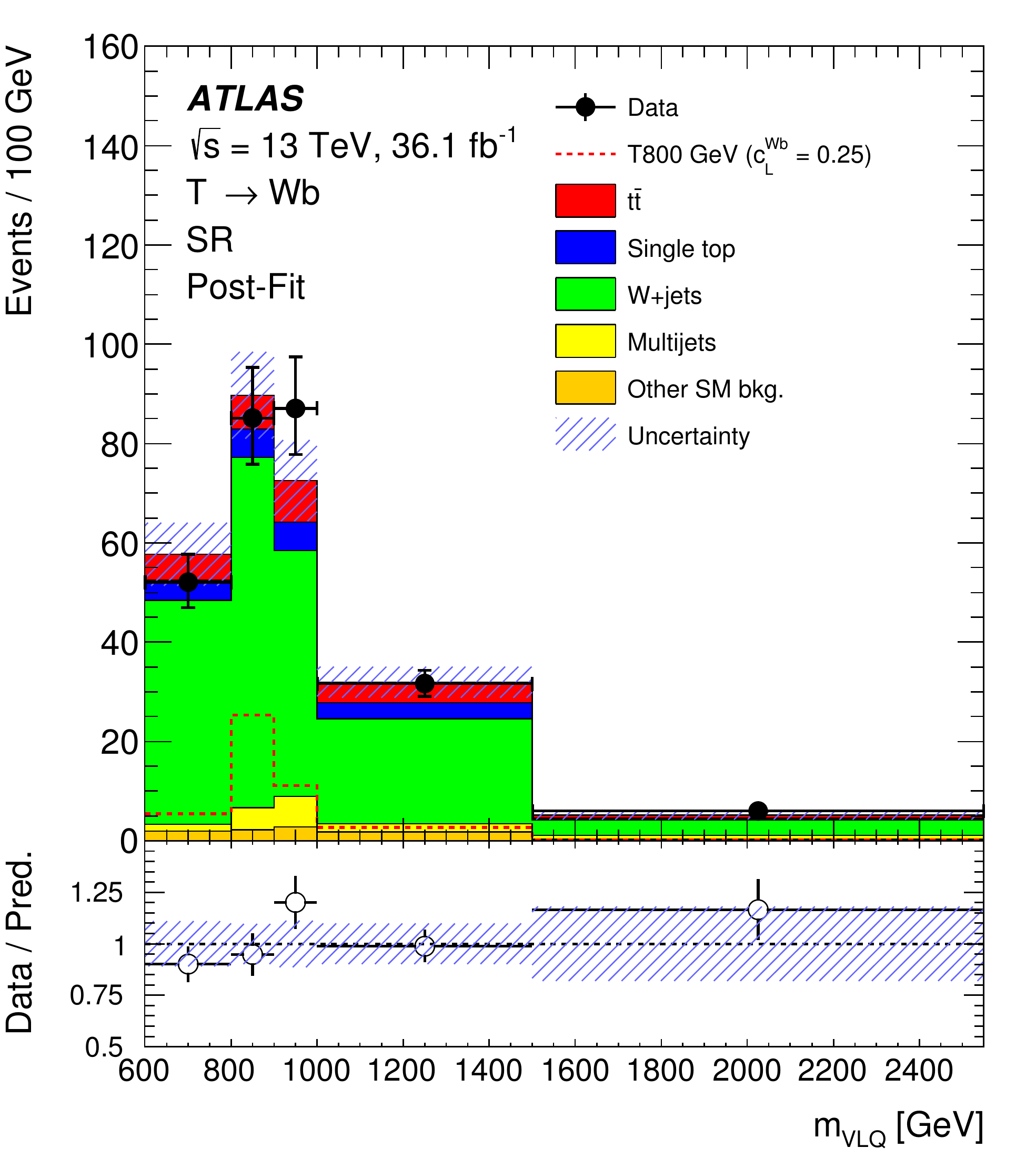}
\label{Fig:VLQ_T1}
}
\hfill
\subfigure[]{
\includegraphics[width=0.45\textwidth]{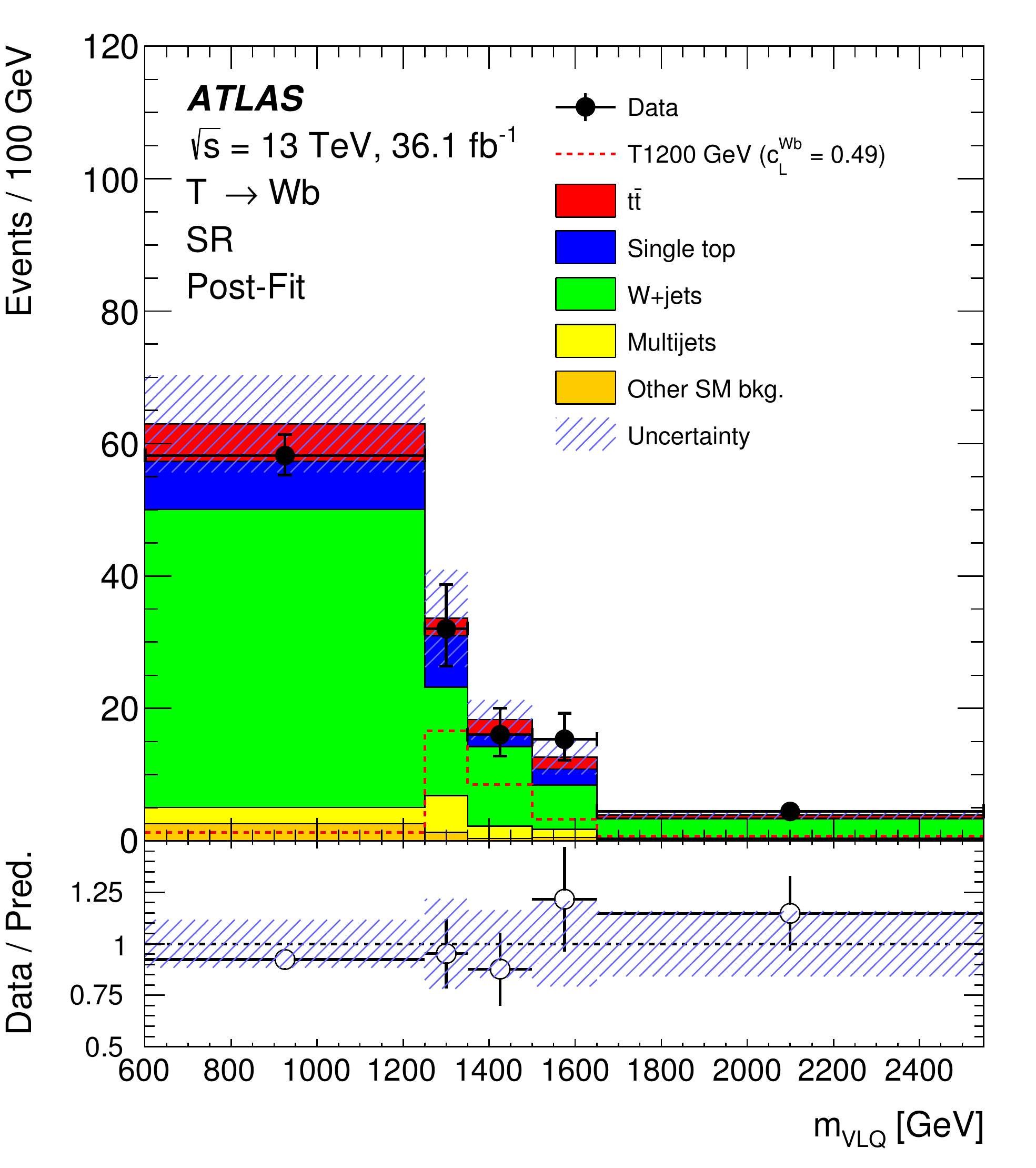}
\label{Fig:VLQ_T2}

}
\null
\caption{Distributions of the VLQ candidate mass, $\mQ$, after the fit to the background-only hypotheses for four different binnings
chosen for four different signal masses. The first and last bin include the underflow
and overflow respectively.
The VLQ candidate mass distributions for \subref{Fig:VLQ_Y1} a left-handed $Y$ signal with mass 900~\GeV\ and coupling $\cL$ = 0.27,  \subref{Fig:VLQ_Y2} a left-handed $Y$ signal with mass 1500~\GeV\ and coupling $\cL$ = 0.64, \subref{Fig:VLQ_T1} a left-handed $T$ signal with mass of 800~\GeV\ and coupling $\cL$ = 0.25 and \subref{Fig:VLQ_T2} a left-handed $T$ signal with mass 1200~\GeV\ and coupling \mbox{$\cL$ = 0.49} are also shown; all signal distributions include interference. The lower panels show the ratio of data to the fitted background yields. The error bars represent the statistical uncertainty in the data. The band represents the total systematic uncertainty after the maximum-likelihood fit.}
\label{fig:VLQ_different_signals}
\end{figure}

Depending on the binning of the $\mQ$ distribution, it is possible that negative entries occur in some bins of the signal-plus-interference template due to the interference effect when large couplings are considered, and this poses a problem in the limit-setting procedure. To avoid this problem, the last bins in the reconstructed $\mQ$ distribution are merged until no negative bin entries exist. As a result, a different binning in the $\mQ$ distribution is chosen for each VLQ mass hypothesis for the $T$-singlet case and for the left-handed $Y$ case, which guarantees (independent of $c^{Wb}$) that all bins in the signal-plus-interference template have positive values. The rebinning reduces the sensitivity for high-mass $T$ and left-handed $Y$ signals.
As an example, Figure~\ref{fig:VLQ_different_signals} shows the fitted VLQ candidate mass distributions for left-handed $Y$ signals with masses of 900~\GeV\ and 1500~\GeV\ and for left-handed $T$ signals with masses of 800~\GeV\ and 1200~\GeV. For the $T$ singlet model, the total integral of the signal-plus-interference template at reconstruction level can become negative for VLQ mass hypotheses above 1200~\GeV. As a result, no coupling-value limits are set for the $T$ singlet model with masses above 1200~\GeV. Tables~\ref{tab:limits_T_all}, \ref{tab:limits_RHY_all}, and \ref{tab:limits_LHY_all} summarise
the observed and expected \mbox{95\% CL}$_\text{s}$ upper limits on
the coupling value and limits on the mixing angle as a function of $Q$-quark mass, for the $T$ singlet model (assuming $\mathcal{B}(T\rightarrow Wb)$ $\approx$ 0.5), the
right-handed $Y$ in a $(B,Y)$ doublet model, and the left-handed $Y$ in a $(T,B,Y)$
triplet model respectively.
The parameterisation of Ref.~\cite{Aguilar-Saavedra:2013qpa} in terms of right- or left-handed mixing angles is chosen
for the coupling limits; these can be easily translated to the
parameterisation of Ref.~\cite{Matsedonskyi:2014mna} for the models under consideration.
In a $T$ singlet model, the upper exclusion limit on $|\sinL|$ ($\cL$) is 0.18 (0.25) for a $T$ quark of mass of 800~\GeV,  rising to 0.35 (0.49) for a $T$ quark with a mass of 1200~\GeV. For a $(B,Y)$ doublet, the upper exclusion limit on $|\sinR|$ ($\cR$) is 0.17 (0.24) for a signal with a mass of 800~\GeV\ and 0.55 (0.77) for $Y$ quarks with a mass of 1800~\GeV. The observed (expected) lower mass limit for $Y$ quarks is about 1.64~\TeV\ (1.80~\TeV) for a right-handed coupling value of $\cR$ = 1/$\sqrt{2}$. For $Y$ signals in a $(T,B,Y)$ triplet the upper exclusion limits on $|\sinL|$ ($\cL$) vary between 0.16 (0.31) and 0.39 (0.78) for masses between 800~\GeV\ and 1600~\GeV.
 
\begin{table}[htb]
\caption{\label{tab:limits_T_all}
Observed and expected 95\% CL upper limits on
$|\sinL|$ and $\cL$ for a left-handed $T$ quark in a $T$ singlet model with masses of 800~\GeV\
to 1200~\GeV\ assuming $\mathcal{B}(T\rightarrow Wb)$ $=$ 0.5. The $\pm 1 \sigma$ and $\pm 2 \sigma$ uncertainties in the expected limits are also given.}
\begin{center}
\renewcommand{\arraystretch}{1.6}
\sisetup{round-mode = places}
\begin{tabular}{rc c  c c}
\toprule\toprule
$T$ mass [\GeV] & Observed limit on   & Expected limit on         &Observed limit on  &Expected limit on \\
& $|\sinL|$ & $|\sinL|$$^{\,+1\sigma/+2\sigma}_{\,-1\sigma/-2\sigma}$    &$\cL$  &$\cL$$^{\, +1\sigma/+2\sigma}_{\, -1\sigma/-2\sigma}$ \\
 
\midrule
800       &  \num[round-precision=2]{0.179322} &\num[round-precision=2]{0.194421} $^{ \num[round-precision=2]{0.039051}/\num[round-precision=2]{0.078489}}_{ \num[round-precision=2]{0.033912}/\num[round-precision=2]{0.055895}}$  &\num[round-precision=2]{0.253600} &\num[round-precision=2]{0.274953}$^{\, \num[round-precision=2]{0.055227}/\num[round-precision=2]{0.111000}}_{\, \num[round-precision=2]{0.047959}/\num[round-precision=2]{0.079047}}$ \\
900       &  \num[round-precision=2]{0.239427} &\num[round-precision=2]{0.204454} $^{ \num[round-precision=2]{0.049227}/\num[round-precision=2]{0.089418}}_{ \num[round-precision=2]{0.047518}/\num[round-precision=2]{0.068994}}$  &\num[round-precision=2]{0.338601} &\num[round-precision=2]{0.289142}$^{\, \num[round-precision=2]{0.069618}/\num[round-precision=2]{0.126456}}_{\, \num[round-precision=2]{0.067201}/\num[round-precision=2]{0.097572}}$  \\
1000      &  \num[round-precision=2]{0.204451} &\num[round-precision=2]{0.212521} $^{ \num[round-precision=2]{0.057146}/\num[round-precision=2]{0.084078}}_{ \num[round-precision=2]{0.067874}/\num[round-precision=2]{0.087675}}$  &\num[round-precision=2]{0.289137} &\num[round-precision=2]{0.300550}$^{\, \num[round-precision=2]{0.080816}/\num[round-precision=2]{0.118904}}_{\, \num[round-precision=2]{0.095989}/\num[round-precision=2]{0.123991}}$ \\
1100      &  \num[round-precision=2]{0.253751} &\num[round-precision=2]{0.269257} $^{ \num[round-precision=2]{0.086246}/\num[round-precision=2]{0.105698}}_{ \num[round-precision=2]{0.125329}/\num[round-precision=2]{0.145021}}$  &\num[round-precision=2]{0.358858} &\num[round-precision=2]{0.380787}$^{\, \num[round-precision=2]{0.121970}/\num[round-precision=2]{0.149479}}_{\, \num[round-precision=2]{0.177242}/\num[round-precision=2]{0.205091}}$ \\
1200      &  \num[round-precision=2]{0.348221} &\num[round-precision=2]{0.349864} $^{ \num[round-precision=2]{0.128348}/\num[round-precision=2]{0.140719}}_{ \num[round-precision=2]{0.216317}/\num[round-precision=2]{0.234583}}$  &\num[round-precision=2]{0.492460} &\num[round-precision=2]{0.494783}$^{\, \num[round-precision=2]{0.181512}/\num[round-precision=2]{0.199007}}_{\, \num[round-precision=2]{0.305918}/\num[round-precision=2]{0.331751}}$ \\
 
\bottomrule\bottomrule
\end{tabular}
\end{center}
\end{table}

\begin{table}[htb]
\begin{center}
\caption{\label{tab:limits_RHY_all}
Observed and expected 95\% CL upper limits on
$|\sinR|$ and $\cR$ for a right-handed $Y$ quark in a $(B,Y)$ doublet model with masses of 800~\GeV\
to 1800~\GeV. The $\pm 1 \sigma$ and $\pm 2 \sigma$ uncertainties in the expected limits are also given.}
\renewcommand{\arraystretch}{1.6}
\sisetup{round-mode = places}
\begin{tabular}{rc c  c c}
\toprule\toprule
$Y$ mass [\GeV] & Observed limit on   & Expected limit on         &Observed limit on  &Expected limit on \\
& $|\sinR|$ & $|\sinR|$$^{\,+1\sigma/+2\sigma}_{\,-1\sigma/-2\sigma}$    &$\cR$  &$\cR$$^{\, +1\sigma/+2\sigma}_{\, -1\sigma/-2\sigma}$ \\
 
\midrule
800       &  \num[round-precision=2]{0.171260} & \num[round-precision=2]{0.195364} $^{ \num[round-precision=2]{0.037229}/ \num[round-precision=2]{0.081533}}_{ \num[round-precision=2]{0.029550}/\num[round-precision=2]{0.052265}}$  &\num[round-precision=2]{0.242198} &\num[round-precision=2]{0.276286}$^{\, \num[round-precision=2]{0.052650}/\num[round-precision=2]{0.115305}}_{\, \num[round-precision=2]{0.041790}/\num[round-precision=2]{0.073913}}$ \\
900       &  \num[round-precision=2]{0.184466} & \num[round-precision=2]{0.188526} $^{ \num[round-precision=2]{0.036262}/ \num[round-precision=2]{0.080702}}_{ \num[round-precision=2]{0.028523}/\num[round-precision=2]{0.050445}}$  &\num[round-precision=2]{0.260875} &\num[round-precision=2]{0.266616}$^{\, \num[round-precision=2]{0.051282}/\num[round-precision=2]{0.114130}}_{\, \num[round-precision=2]{0.040338}/\num[round-precision=2]{0.071340}}$  \\
1000      &  \num[round-precision=2]{0.174145} & \num[round-precision=2]{0.173269} $^{ \num[round-precision=2]{0.031535}/ \num[round-precision=2]{0.067431}}_{ \num[round-precision=2]{0.026188}/\num[round-precision=2]{0.046325}}$  &\num[round-precision=2]{0.246278} &\num[round-precision=2]{0.245040}$^{\, \num[round-precision=2]{0.044597}/\num[round-precision=2]{0.095363}}_{\, \num[round-precision=2]{0.037036}/\num[round-precision=2]{0.065513}}$ \\
1100      &  \num[round-precision=2]{0.167044} & \num[round-precision=2]{0.176555} $^{ \num[round-precision=2]{0.032506}/ \num[round-precision=2]{0.069610}}_{ \num[round-precision=2]{0.026711}/\num[round-precision=2]{0.047238}}$  &\num[round-precision=2]{0.236235} &\num[round-precision=2]{0.249687}$^{\, \num[round-precision=2]{0.045970}/\num[round-precision=2]{0.098443}}_{\, \num[round-precision=2]{0.037775}/\num[round-precision=2]{0.066805}}$ \\
1200      &  \num[round-precision=2]{0.174721} & \num[round-precision=2]{0.200792} $^{ \num[round-precision=2]{0.036918}/ \num[round-precision=2]{0.078878}}_{ \num[round-precision=2]{0.030451}/\num[round-precision=2]{0.053824}}$  &\num[round-precision=2]{0.247092} &\num[round-precision=2]{0.283963}$^{\, \num[round-precision=2]{0.052211}/\num[round-precision=2]{0.111550}}_{\, \num[round-precision=2]{0.043064}/\num[round-precision=2]{0.076119}}$  \\
1300      &  \num[round-precision=2]{0.190795} & \num[round-precision=2]{0.222700} $^{ \num[round-precision=2]{0.040877}/ \num[round-precision=2]{0.087053}}_{ \num[round-precision=2]{0.033830}/\num[round-precision=2]{0.059767}}$  &\num[round-precision=2]{0.269825} &\num[round-precision=2]{0.314945}$^{\, \num[round-precision=2]{0.057809}/\num[round-precision=2]{0.123112}}_{\, \num[round-precision=2]{0.047843}/\num[round-precision=2]{0.084524}}$ \\
1400      &  \num[round-precision=2]{0.244224} & \num[round-precision=2]{0.252401} $^{ \num[round-precision=2]{0.045283}/ \num[round-precision=2]{0.097223}}_{ \num[round-precision=2]{0.038291}/\num[round-precision=2]{0.067634}}$  &\num[round-precision=2]{0.345385} &\num[round-precision=2]{0.356948}$^{\, \num[round-precision=2]{0.064040}/\num[round-precision=2]{0.137495}}_{\, \num[round-precision=2]{0.054152}/\num[round-precision=2]{0.095649}}$  \\
1500      &  \num[round-precision=2]{0.310316} & \num[round-precision=2]{0.276923} $^{ \num[round-precision=2]{0.050380}/ \num[round-precision=2]{0.108895}}_{ \num[round-precision=2]{0.042432}/\num[round-precision=2]{0.074768}}$  &\num[round-precision=2]{0.438853} &\num[round-precision=2]{0.391628}$^{\, \num[round-precision=2]{0.071248}/\num[round-precision=2]{0.154001}}_{\, \num[round-precision=2]{0.060008}/\num[round-precision=2]{0.105738}}$ \\
1600      &  \num[round-precision=2]{0.449691} & \num[round-precision=2]{0.373168} $^{ \num[round-precision=2]{0.079825}/ \num[round-precision=2]{0.189814}}_{ \num[round-precision=2]{0.058363}/\num[round-precision=2]{0.102076}}$  &\num[round-precision=2]{0.635959} &\num[round-precision=2]{0.527740}$^{\, \num[round-precision=2]{0.112890}/\num[round-precision=2]{0.268438}}_{\, \num[round-precision=2]{0.082537}/\num[round-precision=2]{0.144358}}$  \\
1700      &  \num[round-precision=2]{0.587375} & \num[round-precision=2]{0.458969} $^{ \num[round-precision=2]{0.102786}/ \num[round-precision=2]{0.251797}}_{ \num[round-precision=2]{0.075561}/\num[round-precision=2]{0.129990}}$  &\num[round-precision=2]{0.830674} &\num[round-precision=2]{0.649080}$^{\, \num[round-precision=2]{0.145362}/\num[round-precision=2]{0.356095}}_{\, \num[round-precision=2]{0.106859}/\num[round-precision=2]{0.183834}}$ \\
1800      &  \num[round-precision=2]{0.545339} & \num[round-precision=2]{0.432548} $^{ \num[round-precision=2]{0.094860}/ \num[round-precision=2]{0.224957}}_{ \num[round-precision=2]{0.069936}/\num[round-precision=2]{0.120885}}$  &\num[round-precision=2]{0.771225} &\num[round-precision=2]{0.611715}$^{\, \num[round-precision=2]{0.134152}/\num[round-precision=2]{0.318137}}_{\, \num[round-precision=2]{0.098905}/\num[round-precision=2]{0.170958}}$  \\
 
\bottomrule\bottomrule
\end{tabular}
\end{center}
\end{table}

\begin{table}[htb]
\caption{\label{tab:limits_LHY_all}
Observed and expected 95\% CL upper limits on
$|\sinL|$ and $\cL$ for a left-handed $Y$ quark in a $(T,B,Y)$ triplet model with masses of 800~\GeV\
to 1600~\GeV. The $\pm 1 \sigma$ and $\pm 2 \sigma$ uncertainties in the expected limits are also given.}
\begin{center}
\renewcommand{\arraystretch}{1.6}
\sisetup{round-mode = places}
\begin{tabular}{rc c  c c}
\toprule\toprule
$Y$ mass [\GeV] & Observed limit on   & Expected limit on         &Observed limit on  &Expected limit on \\
& $|\sinL|$ & $|\sinL|$$^{\,+1\sigma/+2\sigma}_{\,-1\sigma/-2\sigma}$    &$\cL$  &$\cL$$^{\, +1\sigma/+2\sigma}_{\, -1\sigma/-2\sigma}$ \\
 
\midrule
800       &  \num[round-precision=2]{0.157492}   &\num[round-precision=2]{0.199411} $^{ \num[round-precision=2]{0.040902}/\num[round-precision=2]{0.094057}}_{ \num[round-precision=2]{0.030226}/\num[round-precision=2]{0.053430}}$ &\num[round-precision=2]{0.314983} &\num[round-precision=2]{0.398823} $^{ \num[round-precision=2]{0.081803}/\num[round-precision=2]{0.188113}}_{ \num[round-precision=2]{0.060452}/\num[round-precision=2]{0.106860}}$     \\
900       &  \num[round-precision=2]{0.138233}   &\num[round-precision=2]{0.151247} $^{ \num[round-precision=2]{0.029054}/\num[round-precision=2]{0.066229}}_{ \num[round-precision=2]{0.022895}/\num[round-precision=2]{0.040486}}$ &\num[round-precision=2]{0.276466} &\num[round-precision=2]{0.302494} $^{ \num[round-precision=2]{0.058109}/\num[round-precision=2]{0.132458}}_{ \num[round-precision=2]{0.045791}/\num[round-precision=2]{0.080973}}$ \\
1000      &  \num[round-precision=2]{0.159789}	 &\num[round-precision=2]{0.146162} $^{ \num[round-precision=2]{0.027020}/\num[round-precision=2]{0.058051}}_{ \num[round-precision=2]{0.022099}/\num[round-precision=2]{0.039087}}$ &\num[round-precision=2]{0.319579} &\num[round-precision=2]{0.292323} $^{ \num[round-precision=2]{0.054039}/\num[round-precision=2]{0.116101}}_{ \num[round-precision=2]{0.044199}/\num[round-precision=2]{0.078174}}$ \\
1100      &  \num[round-precision=2]{0.234216}	 &\num[round-precision=2]{0.216594} $^{ \num[round-precision=2]{0.034417}/\num[round-precision=2]{0.075744}}_{ \num[round-precision=2]{0.033064}/\num[round-precision=2]{0.058326}}$ &\num[round-precision=2]{0.468432} &\num[round-precision=2]{0.433187} $^{ \num[round-precision=2]{0.068834}/\num[round-precision=2]{0.151489}}_{ \num[round-precision=2]{0.066128}/\num[round-precision=2]{0.116652}}$\\
1200      &  \num[round-precision=2]{0.201053}	 &\num[round-precision=2]{0.163253} $^{ \num[round-precision=2]{0.030298}/\num[round-precision=2]{0.066252}}_{ \num[round-precision=2]{0.024799}/\num[round-precision=2]{0.043816}}$ &\num[round-precision=2]{0.402107} &\num[round-precision=2]{0.326505} $^{ \num[round-precision=2]{0.060596}/\num[round-precision=2]{0.132504}}_{ \num[round-precision=2]{0.049597}/\num[round-precision=2]{0.087632}}$ \\
1300      &  \num[round-precision=2]{0.246641}	 &\num[round-precision=2]{0.212853} $^{ \num[round-precision=2]{0.037953}/\num[round-precision=2]{0.081508}}_{ \num[round-precision=2]{0.032647}/\num[round-precision=2]{0.057529}}$ &\num[round-precision=2]{0.493282} &\num[round-precision=2]{0.425706} $^{ \num[round-precision=2]{0.075907}/\num[round-precision=2]{0.163016}}_{ \num[round-precision=2]{0.065293}/\num[round-precision=2]{0.115059}}$\\
1400      &  \num[round-precision=2]{0.178464}	 &\num[round-precision=2]{0.254564} $^{ \num[round-precision=2]{0.047225}/\num[round-precision=2]{0.100965}}_{ \num[round-precision=2]{0.039460}/\num[round-precision=2]{0.069259}}$ &\num[round-precision=2]{0.356928} &\num[round-precision=2]{0.509128} $^{ \num[round-precision=2]{0.094450}/\num[round-precision=2]{0.201930}}_{ \num[round-precision=2]{0.078920}/\num[round-precision=2]{0.138518}}$ \\
1500      &  \num[round-precision=2]{0.321251}	 &\num[round-precision=2]{0.348609} $^{ \num[round-precision=2]{0.078153}/\num[round-precision=2]{0.183170}}_{ \num[round-precision=2]{0.058895}/\num[round-precision=2]{0.101143}}$ &\num[round-precision=2]{0.642501} &\num[round-precision=2]{0.697218} $^{ \num[round-precision=2]{0.156306}/\num[round-precision=2]{0.366341}}_{ \num[round-precision=2]{0.117789}/\num[round-precision=2]{0.202287}}$\\
1600      &  \num[round-precision=2]{0.388421}	 &\num[round-precision=2]{0.401102} $^{ \num[round-precision=2]{0.106412}/\num[round-precision=2]{0.280749}}_{ \num[round-precision=2]{0.071866}/\num[round-precision=2]{0.121300}}$ &\num[round-precision=2]{0.776842} &\num[round-precision=2]{0.802204} $^{ \num[round-precision=2]{0.212824}/\num[round-precision=2]{0.561497}}_{ \num[round-precision=2]{0.143732}/\num[round-precision=2]{0.242601}}$ \\

\bottomrule\bottomrule
\end{tabular}
\end{center}
\end{table}

\clearpage
 
\begin{figure}[htb]
\subfigure[]{
\includegraphics[width=0.49\textwidth]{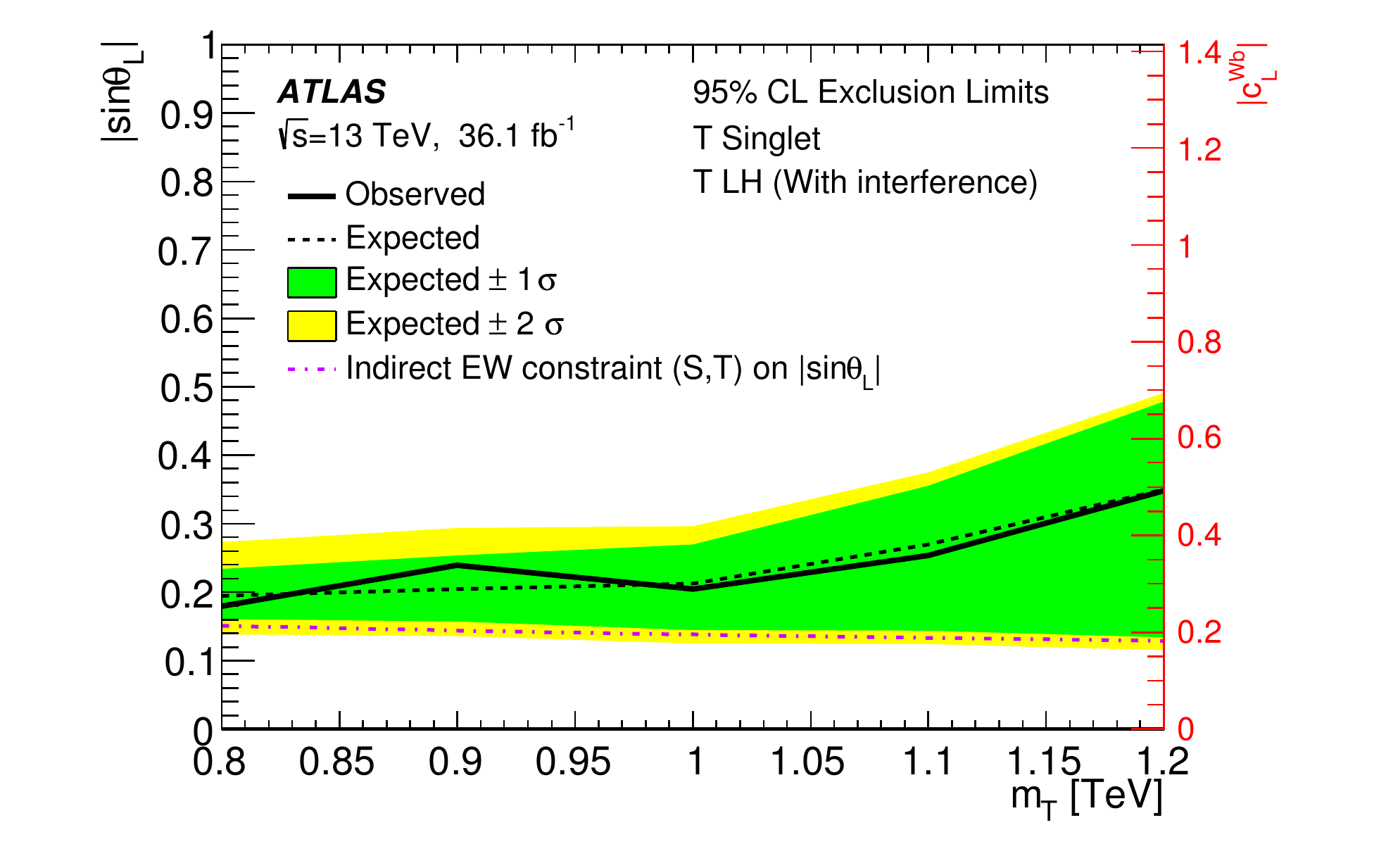}
\label{Fig::mixing_T_with_int}
}
\subfigure[]{
\includegraphics[width=0.49\textwidth]{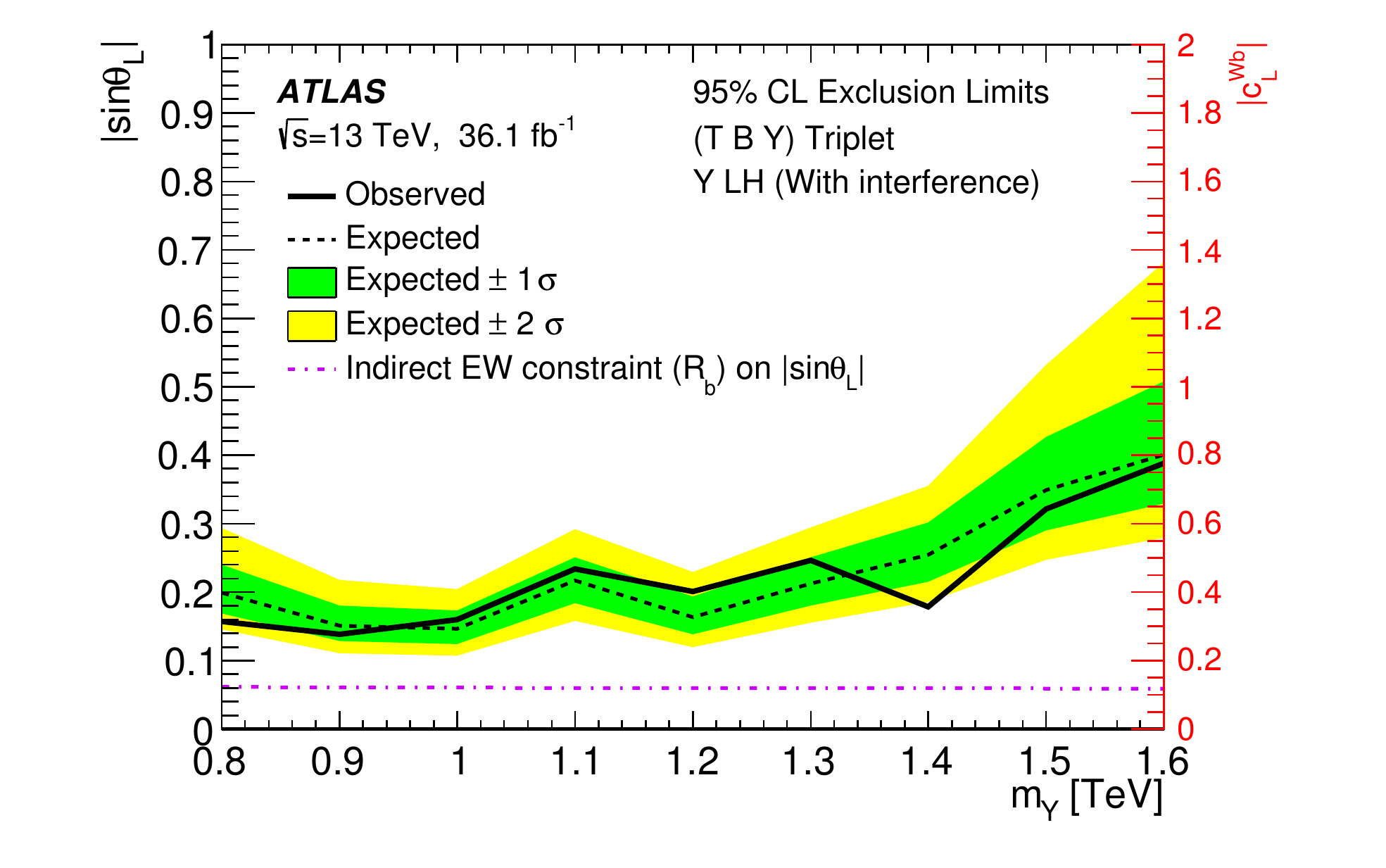}
\label{Fig::mixing_TBY_with_int}
}
\subfigure[]{
\includegraphics[width=0.49\textwidth]{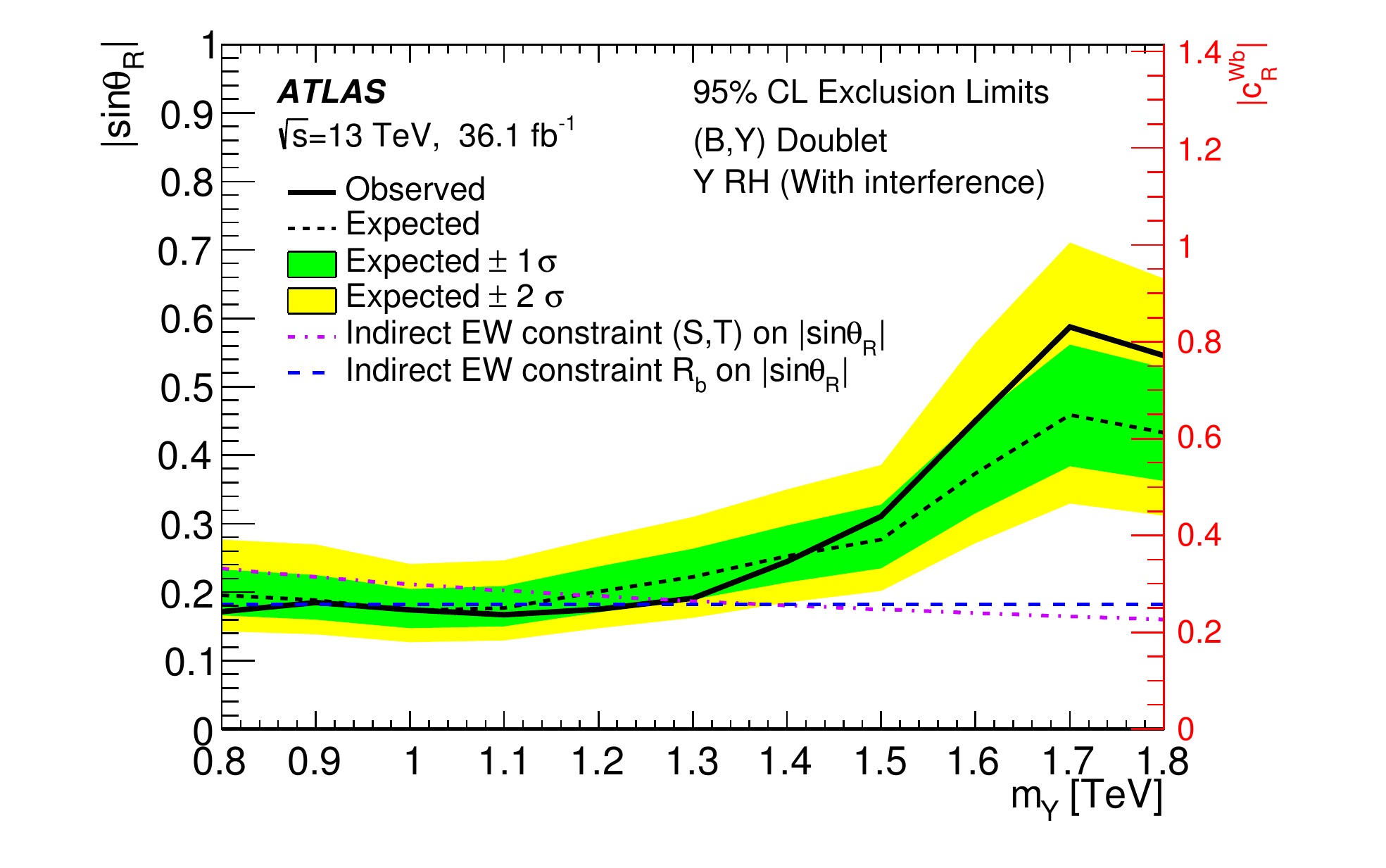}
\label{Fig::mixing_BY_with_int}
}

\caption{\label{Fig::coupling_limits_1_obs_com} Observed (solid line) and expected (short-dashed line) 95$\%$ CL limits on \subref{Fig::mixing_T_with_int} the mixing angle $|\sinL|$ and the coupling value $\cL$ for a singlet
$T$-quark model assuming $\mathcal{B}(T\rightarrow Wb)$ $\approx$ 0.5, \subref{Fig::mixing_TBY_with_int} $|\sinL|$ and $\cL$
for a $(T,B,Y)$ triplet model, and \subref{Fig::mixing_BY_with_int} $|\sinR|$ and $\cR$ for a
$(B,Y)$ doublet model assuming a branching ratio $\mathcal{B}(Y\rightarrow Wb$) = 1, as a function of the VLQ mass. The surrounding bands correspond to $\pm$1 and $\pm$2 standard deviations around the expected limit.
The excluded region is given by the area above the solid line.
Constraints from electroweak precision observables, which are only valid for the mixing angles,
from either the $S$ and $T$ parameters (dashed-dotted line) or from the $R_b$ values (long-dashed line),
are also shown.
These constraints are taken from Ref.~\cite{Aguilar-Saavedra:2013qpa}, where they are presented as a function of
$m_{\mathrm{B}}$ (in the $(B,Y)$ doublet case), respectively, $m_{\mathrm{T}}$ (in the $(T,B,Y)$ triplet case)
and translated to $m_{\mathrm{Y}}$ using the value of the corresponding mixing angle constraint.}
\end{figure}

In Figure~\ref{Fig::coupling_limits_1_obs_com}, these direct mixing-angle bounds are compared with those
from electroweak precision observables taken from Ref.~\cite{Aguilar-Saavedra:2013qpa},
assuming that there are no multiplets other than the one considered. For the $(B,Y)$ doublet model, the bounds presented here are competitive with the indirect constraints for VLQ masses between 800~\GeV\ and 1250~\GeV.
 
\begin{figure}[htb]
\begin{center}
\includegraphics[width=0.65\textwidth]{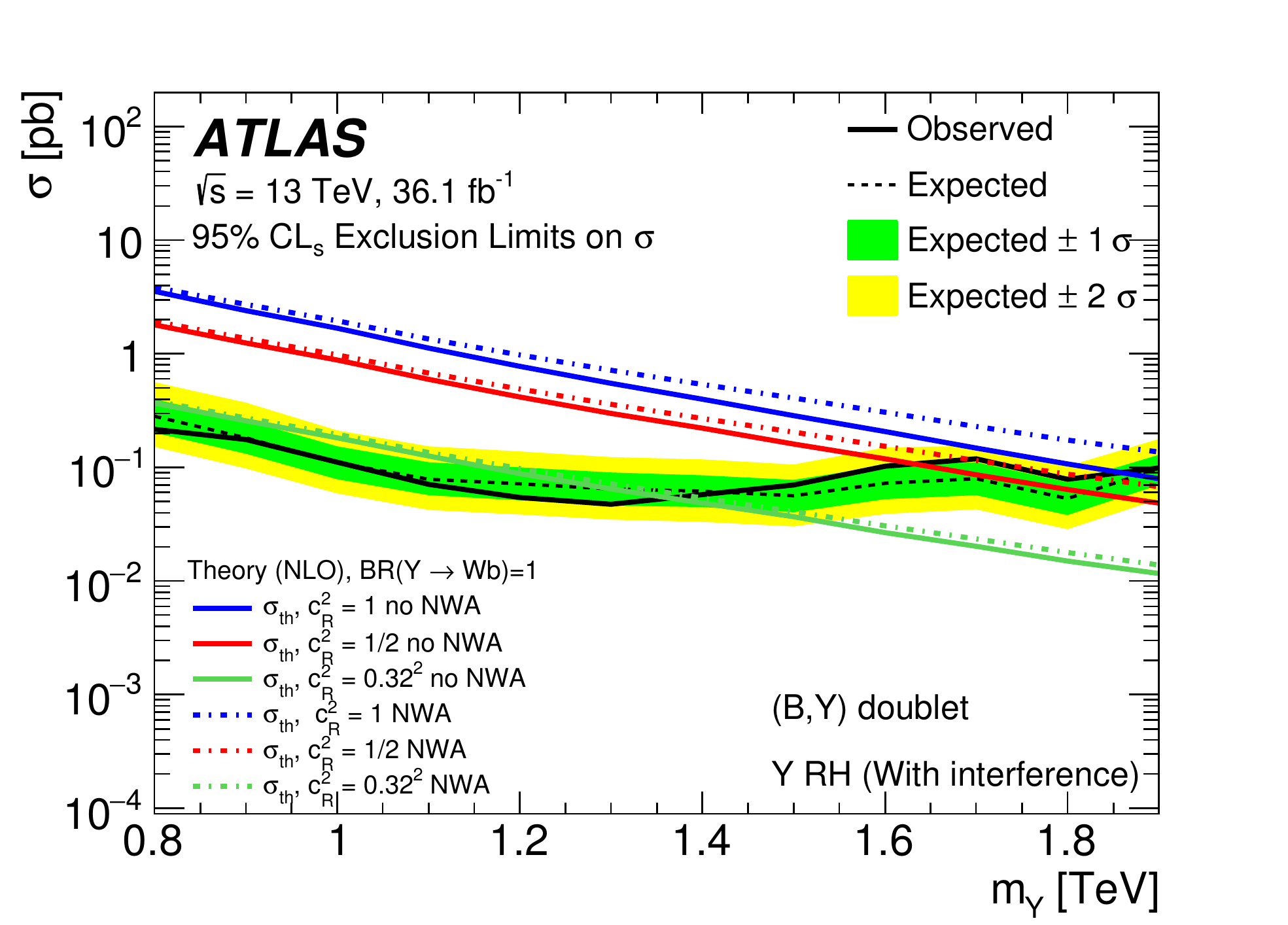}
\caption{Observed (solid line) and expected (short-dashed line) 95$\%$ CL limits on cross-section times branching ratio for the case of the right-handed $Y$ quark for a $(B,Y)$ doublet model as a function of VLQ mass. For the theoretical prediction,
the branching ratio $\mathcal{B}(Y \rightarrow Wb)$ is set to one.
The theoretical NLO cross-sections for different coupling values are shown for the
calculation using the narrow-width approximation (dashed-dotted lines) and using
no narrow-width approximation (solid lines) as described in the text.
\hfill
\label{fig:cross-section_times_BR_RHY}}
\end{center}
\end{figure}

Since the interference effect for the case of the right-handed $Y$ quark is very small,
and therefore the signal+interference template is very similar to the one of a pure resonance,
a limit on $\sigma_{\text {VLQ}}+\sigma_{\text {I}}$ times branching ratio is presented
for this case in Figure~\ref{fig:cross-section_times_BR_RHY}, corresponding to the $|\sinR|$
and $\cR$ limits for a $(B,Y)$ doublet model presented in Figure~\subref{Fig::mixing_BY_with_int}.
\clearpage

\section{Conclusion}
\label{sec:conclusion}
A search for the production of a single vector-like quark $Q$, where $Q$ can be either a $T$ or $Y$ quark,
with the subsequent decay into $Wb$ has been carried out with the ATLAS experiment at the CERN LHC.
The data used in this search correspond to an integrated luminosity of $36.1~\ifb$ of $pp$ collisions with a centre-of-mass energy $\sqrt{s}$ = 13~\TeV\ recorded in 2015 and 2016.
The selected events have exactly one isolated electron or muon, a high-\pT\ $b$-tagged jet,
missing transverse momentum and at least one forward jet. The $Q$ candidate is fully reconstructed
and its mass is used as a discriminating variable in a maximum-likelihood fit.
The observed data distributions are compatible with the expected Standard Model background and no significant excess is observed.
The search result is interpreted for $Q=T$ in a $T$ singlet model and $Q=Y$ in either a $(B,Y)$ doublet model
or in a $(T,B,Y)$ triplet model, taking into account the interference effect with the Standard Model background.
Limits at 95\% CL are set on the cross-section times branching ratio as a function of the VLQ mass
in the case of the $(B,Y)$ doublet model, where interference has the smallest effect. The search results are translated
into limits on the $QWb$ mixing angle or coupling. In the $T$-singlet case, the 95$\%$ CL
limit on $|\sinL|$ ($\cL$) varies between 0.18 and 0.35 (0.25 and 0.49) for masses from
800~\GeV\ to 1200~\GeV. In the $(B,Y)$ doublet model, exclusion limits on $|\sinR|$ ($\cR$) vary between 0.17 and 0.55 (0.24 and 0.77) for masses between 800~\GeV\ and 1800~\GeV\ and the $|\sinR|$ bounds presented here are below the indirect electroweak constraints for masses between about 900~\GeV\ and 1250~\GeV\ where exclusion limits on $|\sinR|$ are around 0.18--0.19. In the case of the $(T,Y,B)$ triplet, the limits on $|\sinL|$ ($\cL$) vary between 0.16 and 0.39 (0.31 and 0.78) for masses from 800~\GeV\ to 1600~\GeV. For all signal scenarios explored, this analysis is found to significantly improve upon the reach of previous searches.
 
\clearpage
 
\section*{Acknowledgements}
 
We thank CERN for the very successful operation of the LHC, as well as the
support staff from our institutions without whom ATLAS could not be
operated efficiently.
 
We acknowledge the support of ANPCyT, Argentina; YerPhI, Armenia; ARC, Australia; BMWFW and FWF, Austria; ANAS, Azerbaijan; SSTC, Belarus; CNPq and FAPESP, Brazil; NSERC, NRC and CFI, Canada; CERN; CONICYT, Chile; CAS, MOST and NSFC, China; COLCIENCIAS, Colombia; MSMT CR, MPO CR and VSC CR, Czech Republic; DNRF and DNSRC, Denmark; IN2P3-CNRS, CEA-DRF/IRFU, France; SRNSFG, Georgia; BMBF, HGF, and MPG, Germany; GSRT, Greece; RGC, Hong Kong SAR, China; ISF and Benoziyo Center, Israel; INFN, Italy; MEXT and JSPS, Japan; CNRST, Morocco; NWO, Netherlands; RCN, Norway; MNiSW and NCN, Poland; FCT, Portugal; MNE/IFA, Romania; MES of Russia and NRC KI, Russian Federation; JINR; MESTD, Serbia; MSSR, Slovakia; ARRS and MIZ\v{S}, Slovenia; DST/NRF, South Africa; MINECO, Spain; SRC and Wallenberg Foundation, Sweden; SERI, SNSF and Cantons of Bern and Geneva, Switzerland; MOST, Taiwan; TAEK, Turkey; STFC, United Kingdom; DOE and NSF, United States of America. In addition, individual groups and members have received support from BCKDF, CANARIE, CRC and Compute Canada, Canada; COST, ERC, ERDF, Horizon 2020, and Marie Sk{\l}odowska-Curie Actions, European Union; Investissements d' Avenir Labex and Idex, ANR, France; DFG and AvH Foundation, Germany; Herakleitos, Thales and Aristeia programmes co-financed by EU-ESF and the Greek NSRF, Greece; BSF-NSF and GIF, Israel; CERCA Programme Generalitat de Catalunya, Spain; The Royal Society and Leverhulme Trust, United Kingdom.
 
The crucial computing support from all WLCG partners is acknowledged gratefully, in particular from CERN, the ATLAS Tier-1 facilities at TRIUMF (Canada), NDGF (Denmark, Norway, Sweden), CC-IN2P3 (France), KIT/GridKA (Germany), INFN-CNAF (Italy), NL-T1 (Netherlands), PIC (Spain), ASGC (Taiwan), RAL (UK) and BNL (USA), the Tier-2 facilities worldwide and large non-WLCG resource providers. Major contributors of computing resources are listed in Ref.~\cite{ATL-GEN-PUB-2016-002}.
 
 
\printbibliography
 
\clearpage
 
\begin{flushleft}
{\Large The ATLAS Collaboration}

\bigskip

M.~Aaboud$^\textrm{\scriptsize 34d}$,    
G.~Aad$^\textrm{\scriptsize 99}$,    
B.~Abbott$^\textrm{\scriptsize 124}$,    
O.~Abdinov$^\textrm{\scriptsize 13,*}$,    
B.~Abeloos$^\textrm{\scriptsize 128}$,    
D.K.~Abhayasinghe$^\textrm{\scriptsize 91}$,    
S.H.~Abidi$^\textrm{\scriptsize 164}$,    
O.S.~AbouZeid$^\textrm{\scriptsize 39}$,    
N.L.~Abraham$^\textrm{\scriptsize 153}$,    
H.~Abramowicz$^\textrm{\scriptsize 158}$
% ...
\bigskip
\\

% ...
$^{13}$Institute of Physics, Azerbaijan Academy of Sciences, Baku; Azerbaijan.\\
% ...
$^{34}$$^{(a)}$Facult\'e des Sciences Ain Chock, R\'eseau Universitaire de Physique des Hautes Energies - Universit\'e Hassan II, Casablanca;$^{(b)}$Centre National de l'Energie des Sciences Techniques Nucleaires (CNESTEN), Rabat;$^{(c)}$Facult\'e des Sciences Semlalia, Universit\'e Cadi Ayyad, LPHEA-Marrakech;$^{(d)}$Facult\'e des Sciences, Universit\'e Mohamed Premier and LPTPM, Oujda;$^{(e)}$Facult\'e des sciences, Universit\'e Mohammed V, Rabat; Morocco.\\
% ...
$^{39}$Niels Bohr Institute, University of Copenhagen, Copenhagen; Denmark.\\
% ...
$^{91}$Department of Physics, Royal Holloway University of London, Egham; United Kingdom.\\
% ...
$^{99}$CPPM, Aix-Marseille Universit\'e, CNRS/IN2P3, Marseille; France.\\
% ...
$^{124}$Homer L. Dodge Department of Physics and Astronomy, University of Oklahoma, Norman OK; United States of America.\\
% ...
$^{128}$LAL, Universit\'e Paris-Sud, CNRS/IN2P3, Universit\'e Paris-Saclay, Orsay; France.\\
% ...
$^{153}$Department of Physics and Astronomy, University of Sussex, Brighton; United Kingdom.\\
% ...
$^{158}$Raymond and Beverly Sackler School of Physics and Astronomy, Tel Aviv University, Tel Aviv; Israel.\\
% ...
$^{164}$Department of Physics, University of Toronto, Toronto ON; Canada.\\
% ...
$^{a}$ Also at Borough of Manhattan Community College, City University of New York, NY; United States of America.\\
$^{b}$ Also at Centre for High Performance Computing, CSIR Campus, Rosebank, Cape Town; South Africa.\\
$^{c}$ Also at CERN, Geneva; Switzerland.\\
$^{d}$ Also at CPPM, Aix-Marseille Universit\'e, CNRS/IN2P3, Marseille; France.\\
$^{e}$ Also at D\'epartement de Physique Nucl\'eaire et Corpusculaire, Universit\'e de Gen\`eve, Gen\`eve; Switzerland.\\
% ...
$^{*}$ Deceased

\end{flushleft}


\end{document}